\documentclass[iop,apj]{emulateapj}
\usepackage{amsmath,amssymb,amstext}

\usepackage[breaklinks,colorlinks,citecolor=blue,linkcolor=magenta]{hyperref} 

\usepackage[all]{hypcap} 

\usepackage{aas_macros}
\usepackage{natbib}
\shorttitle{A Versatile Family of Galactic Wind Models}
\shortauthors{Bustard et al.}

\begin{document}

\title{A Versatile Family of Galactic Wind Models}

\author{
Chad Bustard\altaffilmark{1}
Ellen G. Zweibel\altaffilmark{1,2}
Elena D'Onghia\altaffilmark{2,3}
}
\altaffiltext{1}{Physics Department, University of Wisconsin-Madison, 1150 University Avenue, Madison, WI 53706; bustard@wisc.edu}
\altaffiltext{2}{Department of Astronomy, University of Wisconsin-Madison, 2535 Sterling Hall, 475 N. Charter Street, Madison, WI 53706}
\altaffiltext{3}{Alfred P. Sloan Fellow}

\begin{abstract}
We present a versatile family of model galactic outflows including non-uniform mass and energy source distributions, a gravitational potential from an extended mass source, and radiative losses. The model easily produces steady-state wind solutions for a range of mass-loading factors, energy-loading factors, galaxy mass and galaxy radius. We find that, with radiative losses included, highly mass-loaded winds must be driven at high central temperatures, whereas low mass-loaded winds can be driven at low temperatures just above the peak of the cooling curve, meaning radiative losses can drastically affect the wind solution even for low mass-loading factors. By including radiative losses, we are able to show that subsonic flows can be ignored as a possible mechanism for expelling mass and energy from a galaxy compared to the more efficient transonic solutions. Specifically, the transonic solutions with low mass-loading and high energy-loading are the most efficient. Our model also produces low-temperature, high-velocity winds that could explain the prevalence of low-temperature material in observed outflows. Finally, we show that our model, unlike the well-known \cite{1985Natur.317...44C} model, can reproduce the observed linear relationship between wind X-ray luminosity and star formation rate (SFR) over a large range of SFR from $1-1000$ M$_{\odot}$/yr assuming the wind mass-loading factor is higher for low-mass, and hence, low-SFR galaxies. We also constrain the allowed mass-loading factors that can fit the observed X-ray luminosity vs. SFR trend, further suggesting an inverse relationship between mass-loading and SFR as explored in advanced numerical simulations. 
\end{abstract}

\keywords{galaxies: evolution --- galaxies: fundamental parameters --- galaxies: star formation --- radiation: dynamics --- X-rays: galaxies}
\maketitle

\clearpage
\section{Introduction}
Due to the large variety of physical phenomena occurring on a vast range of scales, galaxy formation
is very challenging to model theoretically, and the full complexity of many processes still needs to be unraveled. Currently, many models of galaxy formation without some form of feedback form stars too early and too quickly, consequently overestimating the number of low-mass galaxies and their stellar ages compared to observations (\cite{1991ApJ...379...52W}, \cite{2006MNRAS.373.1265O}, \cite{2011MNRAS.410.2625P}, \cite{2012MNRAS.421.3522H}, \cite{2013MNRAS.436.3031V}). One possible solution is to introduce a more efficient galactic wind feedback model in these systems. Galactic winds from supernova rich starburst regions and active galactic nuclei transfer mass and energy away from the regions of wind development and enrich the intergalactic medium with metals, eventually suppressing star formation in the galaxies. 

There is a wealth of observational evidence for galactic winds (\cite{2003ApJ...594..695P}, \cite{2003ApJ...588...65S}, \cite{2005ApJ...621..227M}, \cite{2005ARA&A..43..769V}, \cite{2010ApJ...717..289S}, \cite{2011ApJ...733..101G}). However, large-scale galactic outflows, both their intrinsic properties and their effects on galaxy formation and evolution, are poorly understood. Simulations of thermally driven winds (\cite{1985Natur.317...44C}, \cite{1991ApJ...376..380C}, \cite{2015ApJ...814....4L}), radiation driven winds (\cite{2005ApJ...618..569M}, \cite{2011ApJ...735...66M}), and cosmic ray driven winds (\cite{2008ApJ...674..258E}, \cite{2010ApJ...711...13E}, \cite{2012MNRAS.423.2374U}, \cite{2013ApJ...777L..38H}, \cite{2014MNRAS.437.3312S}) have been developed and successfully reproduce observations of galaxies, including the Milky Way and the well-known starburst galaxy, M82. However, these physical processes are at too small scales to be properly accounted for in cosmological simulations. Instead, hydrodynamical simulations of galaxy formation (\cite{2010MNRAS.402.1536S}, \cite{2012MNRAS.425.1270S}, \cite{2012MNRAS.425.3024V}, \cite{2013MNRAS.434.3142A}, \cite{2013MNRAS.430.1901H}, \cite{2015arXiv150800007C}, \cite{2015arXiv150103155M}, \cite{2015MNRAS.446..521S}) typically capture these feedback processes using sub-resolution physics guided by parameterizations from observations. These outflow prescriptions cannot be specific to one or a few galaxies; instead, they must be general enough to encompass galaxies on many scales. It is logical, then, to propose simplified but versatile models for galactic winds and attempt, with those models, to determine what the most important parameters are for driving a wind and how these simplified models compare to observations.

In Section \ref{Model}, we describe our complete re-working of the \cite{1985Natur.317...44C} model, including non-uniform mass and energy source distributions, a gravitational potential from an extended mass distribution, and radiative losses. In Section \ref{critPoints}, we solve for a set of critical points at which the flow transitions from subsonic to supersonic. We do this after scaling the system of steady-state wind equations by $GM/R$, as shown in Section \ref{Scaling}, and using the logical mass-loading and energy-loading parameterizations of Section \ref{Parameterizations}. We show the critical points for a number of different mass-loading and energy-loading factors in Section \ref{determineCritPoints}. In Section \ref{radiative}, we describe the effects of radiative losses on the wind solutions. We find that radiative losses can be important for all ranges of mass-loading factors, even the least mass-loaded winds, which can generally be driven at lower temperatures near the peak of the cooling curve. We also show in Section \ref{efficiency} that radiative losses can severely decrease the efficiency in which winds expel mass from a galaxy. For subsonic winds, as shown in Section \ref{subsonic}, radiative losses sharply decrease the temperature, and hence energy per mass expelled in the wind, at a radius dependent on the energy-loading factor. For transonic solutions, we show in Section \ref{efficiency1} that the least mass-loaded outflows with the highest energy-loading factors are the most efficient. In Section \ref{lowTemp}, we also note the existence of very low temperature, high velocity transonic outflows in our model due to radiative losses. Finally, we compare, in Section \ref{X-ray} the resulting relationship between outflow X-ray luminosity and star formation rate (SFR) to recent observations. Unlike the classical Chevalier and Clegg model, which has been shown to produce a quadratic relationship between luminosity and star formation rate (\cite{2014ApJ...784...93Z}), including gravity and supposing higher mass-loading fractions for lower star formation rates can fairly well produce the observed linear relationship. Specifically, we show that, to match the linear behavior of the observations, the allowed mass-loading factors generally must decrease monotonically as star formation rate, and hence galaxy mass, increases. 

\section{Model}
\label{Model}
One of the standard models for galactic winds was proposed by \cite{1985Natur.317...44C}, hereafter referred to as the CC model. The model supposes a spherically symmetric wind, does not include a gravitational potential, and supposes a constant mass injection $q = \frac{\dot{M}}{V}$  and constant energy injection $Q = \frac{\dot{E}}{V}$ out to a radius R. For $r > R$, $q = Q = 0$. For a spherical geometry, the mass conservation, momentum, and energy equations are
 \begin{equation}
 \label{massEQN}
 \frac{1}{r^2}\frac{\partial}{\partial r}r^2\rho u = q(r)
 \end{equation}
 \begin{equation}
 \label{momEQN}
 \rho u \frac{\partial u}{\partial r} = -qu - \frac{\partial P}{\partial r} - \rho \frac{\partial \Phi}{\partial r}
 \end{equation}
 \begin{equation}
 \label{CCEnergy}
 \frac{1}{r^{2}} \frac{\partial}{\partial r} \left[ \rho u r^{2} \left(\frac{1}{2}u^{2} + \frac{\gamma}{\gamma - 1} \frac{P}{\rho} \right) \right ]= Q_{cc}
 \end{equation}
 where u is the wind velocity. For the CC model, $q(r) = q_{0} = \text{constant}$ for $r < R$, and $d\Phi/dr = 0$. This model was originally proposed to describe M82, a well-studied starburst galaxy with an extremely strong wind. X-ray fits indicate high terminal wind speeds of up to $\approx 1400-2200$ km/s (\cite{2009ApJ...697.2030S}), so far above the escape velocity of the galaxy that there is no need to include gravity in a model to describe M82; hence, the CC model omits it. 

\cite{1958ApJ...128..664P} showed that there are infinitely many wind solutions, all corresponding to different boundary conditions. Many of these are ``breeze'' solutions where the wind accelerates up until the ``Parker point'' and then decelerates to zero velocity again, always staying subsonic. We discuss subsonic outflows in Section \ref{subsonic}, but we show that radiative losses inhibit their efficiency, limiting the mass flux to certain radii before a sharp drop-off in temperature. The most effective type of wind for transferring matter out of a galaxy is the transonic solution that monotonically increases in velocity, becomes supersonic at a critical radius $r_{c}$, and has an asymptotic velocity considerably greater than the subsonic solutions. For now, we just consider transonic solutions.

One can solve the CC mass conservation, momentum, and energy equations for the Mach number of this transonic wind as a function of radius. The radius, R, at which there is a discontinuity in mass and energy input is also the critical radius, $r_{c}$ for the CC model, i.e. it is the radius at which the wind switches continuously from subsonic at $ r < R$ to supersonic at $r > R$. A noticeable, non-physical consequence of this model is that the velocity derivative is infinite at radius R. This is a consequence of the jump in mass and energy injection at that radius. 

\begin{figure}[ht]
\label{CCCompPlot_log2}
\centering
\includegraphics[scale = 0.7]{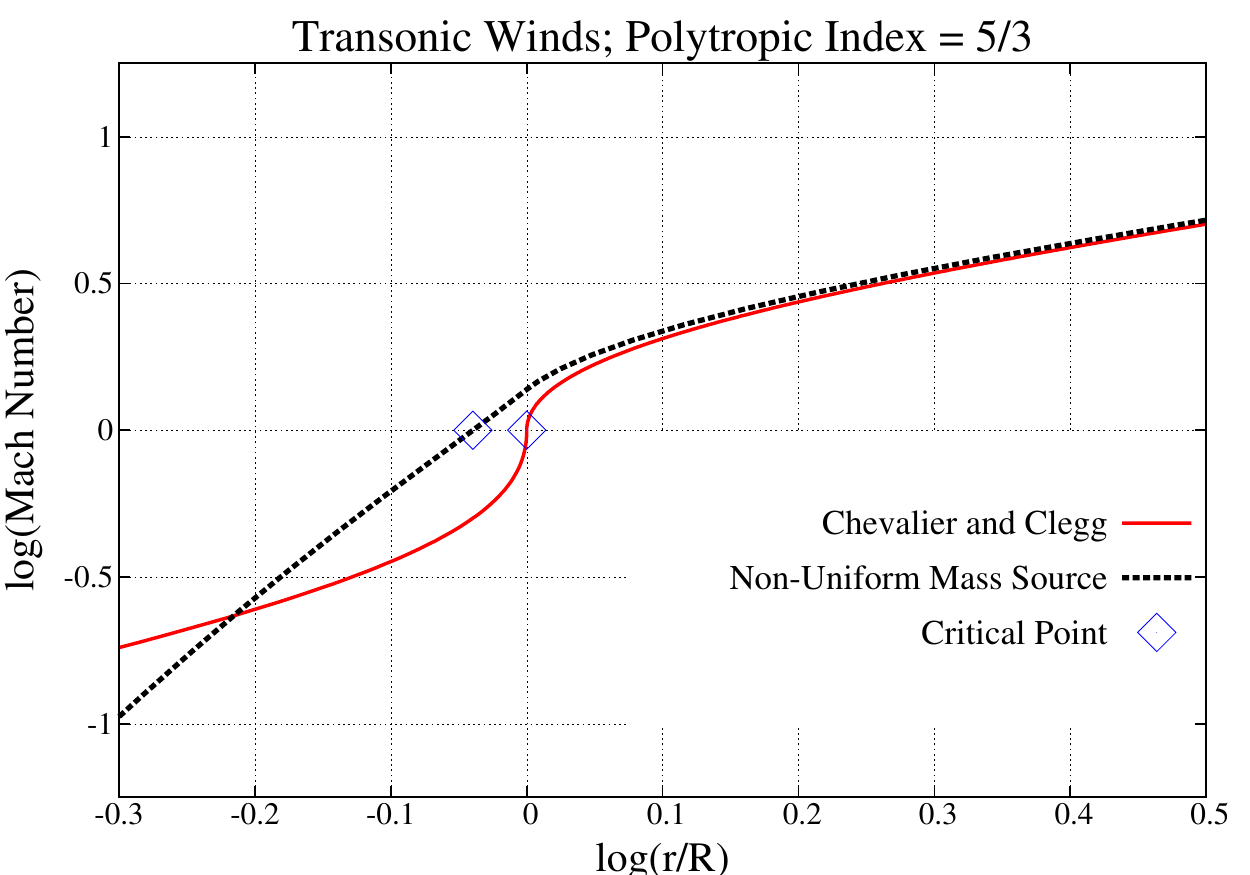}
\caption{The CC model is not analytic at the critical radius R. This
problem stems from having a constant mass source $q_{0}$. Using a mass source such
as $q(r) = q_{0}(1-\frac{r^{2}}{R^{2}})$ gives a finite slope at the now shifted critical point.}
\end{figure}

To instead get an analytic velocity solution at $r = R$, we modify the mass injection to now be a function of r, namely 
\begin{equation}
q(r) = q_{0}(1-\frac{r^{2}}{R^{2}})
\end{equation}
out to a radius R.  We choose this idealized mass injection profile instead of, e.g. an exponential or power law profile, because it allows for tractable, analytic calculations while still retaining the logical and observed behavior that more mass per volume should be injected to the wind near the center of the galaxy. The result of using a non-uniform mass source instead of the CC constant mass source is shown in Fig. \ref{CCCompPlot_log2} for a polytropic wind with polytropic index $\gamma = 5/3$ in the absence of gravity and energy injection. Note that including a non-uniform mass source shifts the critical point so it is no longer exactly at $r = R$. This will prove to be important, as there can actually be infinitely many critical points corresponding to solutions with different central conditions. For each polytropic wind, such as that shown in Fig. \ref{CCCompPlot_log2}, the shifted critical point for our q(r) profile is always very close to $r/R = 1$ regardless of the factor K in the wind equation of state $P = K\rho^{\gamma}$. In addition, in the region $r > R$ where there is no mass or energy addition, the wind expands exactly as it does for the CC model. One can see in Fig. \ref{CCCompPlot_log2} that the asymptotic behavior for each model is very similar. Therefore, we do not believe that the results in this paper are sensitively dependent on the q(r) profile we have chosen.

It should also be noted that the CC model energy equation, eqn. (\ref{CCEnergy}), is not the same as the one we use in our own model, though it is very similar. In the CC model, each term on the left hand side has units of energy per time per volume, hence it is logical to evolve those quantities with an energy per time per volume factor, $Q_{cc}$. 

We choose instead to determine the temperature evolution by the first law of thermodynamics; $dU=dQ+Pd\rho/\rho^2$, where $dQ=TdS$, and $Q$ is not to be confused with the CC source term in their energy equation. Using $P=(\gamma - 1)\rho U$ and assuming a steady state with radial flow $u$, the temperature equation is
\begin{equation}
\label{myTEQN}
u\frac{k_B}{m}\frac{dT}{dr}=(\gamma - 1)\frac{dQ}{dt}+(\gamma - 1)\frac{k_BTu}{m\rho}\frac{d\rho}{dr}
\end{equation}

and the pressure is described by

\begin{equation}
\label{myPEQN}
\frac{dP}{dr}=c_s^2\frac{d\rho}{dr}+\frac{(\gamma - 1)\rho}{u}\frac{dQ}{dt}
\end{equation}
where $c_s^2\equiv\gamma k_BT/m$ is the sound speed squared, and we set $\gamma = 5/3$ in all cases studied in this paper. If $dQ/dt\equiv 0$, i.e. we have no additional energy injection, then the resulting flow is adiabatic. The CC model, on the other hand, requires this extra energy injection to drive a wind because the central pressure would be zero without energy addition, whereas our model always has a non-zero pressure at the galactic center. 

A logical way to shape the additional energy addition term, $dQ/dt$, is to give it a $1-\frac{r^{2}}{R^{2}}$ profile, as is the case for the mass injection. Therefore, we define an energy per mass parameter $\eta$ such that 
\begin{equation}
\rho \frac{\partial Q}{\partial t} = \eta q 
\end{equation}

Equation (\ref{massEQN}) is solved in the Appendix, giving 
\begin{equation}
\label{rho2}
\rho=\frac{r\langle q\rangle}{3u},
\end{equation}
where $\langle q\rangle$ is the mean $q$ inside radius $r$. 

We also extend the model to include an extended mass distribution chosen for simplicity to have constant mass density 

\begin{align}
\rho_{m} = 3M/(4\pi R_{m}^{3}) \qquad &  \text{for } r < R_{m}
\end{align}

While this is not fully realistic, having the star-forming mass more centrally concentrated than the total mass replicates some features of central starbursts. Inclusion of a non-uniform mass density with an extended halo is the subject of current work.

Our gravitational potential is then defined in two regions: 
\begin{align}
    \Phi = \frac{GM}{2R_{m}}(\frac{r^{2}}{R_{m}^{2}} - 3) \qquad & \text{for } r < R_{m} \\
    \Phi = -\frac{GM}{r} \qquad  & \text{for } r > R_{m}
\end{align}
In this paper, we always set $R_{m} = R$, i.e. we tie the end of the mass distribution to the end of the mass and energy source distribution; however, the sonic point is not restricted to this radius, as we will see in Section \ref{determineCritPoints}.

\section{Methodology}
\label{critPoints}
\subsection{Scaling by $\frac{GM}{R}$}
\label{Scaling}
It is useful to scale the main wind equations by the quantity $\frac{GM}{R}$, which represents the magnitude of the gravitational potential energy per unit mass at $r = R$. By doing so, each of our solutions to the scaled equations will in fact give us a multitude of solutions for various gravitational potentials. In addition, we write our equations in terms of the dimensionless radius $\tilde{r} = r/R$ (see eqns. (\ref{radU}) and (\ref{radCS}) for the momentum and sound speed equations, respectively, in terms of scaled variables). The resulting scaling for the main wind quantities is then

\begin{equation}
\rho = \tilde{\rho}\sqrt{\frac{R^{3}}{GM}}
\end{equation} 
\begin{equation}
c_{s} = \tilde{c_{s}} \sqrt{GM/R}
\end{equation}
\begin{equation}
u = \tilde{u} \sqrt{GM/R}
\end{equation}
where the tilde denotes the scaled variable. The energy per mass addition also scales accordingly: $\eta = \tilde{\eta} GM/R$. 

\subsection{Parameterizations}
\label{Parameterizations}
To include mass-loading in our modified CC model, we parameterize the mass-loading efficiency by 
\begin{equation}
\dot{M} = \beta  \text{SFR} (\text{M}_{\odot}\text{/yr})
\end{equation}
Then our mass-loading per volume factor, $q_{0}$, is calculated such that $\dot{M} = \int_{0}^{R} q_{0}(1-r^{2}/R^{2}) dV$.
\begin{equation}
q_{0} = \frac{\beta \text{SFR}}{\frac{8}{15}\pi R^{3}} = 1.60 \times 10^{-37} \beta \text{SFR}
\end{equation}
fixing $R = 200$ pc for each galaxy, regardless of the galaxy's mass and SFR. Then our gravitational potential is calculated as $GM/R$. We consider galaxies of dynamical mass $10^{8} - 10^{12} \text{M}_{\odot}$, and the assumption that this entire mass is confined to a radius of 200 pc from the center is quite unreasonable for many galaxies. We are, however, considering only starburst galaxies, for which $10^{7}-10^{8} \text{M}_{\odot}$ of molecular gas can be present near the galactic center. In future simulations, we will change the galaxy radius and see what effect that has on our wind solutions. We would also like to study the effect of a more realistic mass distribution, e.g. including a dark matter halo, etc. Including even a simple gravitational potential, which is no worse than including a point mass potential, allows us to model the many galaxies that do not have a very strong, M82-like wind. For these galaxies, the inward pull of gravity is quite important, regardless of whether the gravitational potential is strictly realistic. 

We parameterize energy addition as 
\begin{equation}
\dot{E} = \alpha' \dot{E}_{SN}
\end{equation}

$\dot{E}_{SN} = \epsilon \nu \text{SFR}$ is the energy per time injected into the wind from supernovae where $\epsilon = 10^{51} \epsilon_{51}$ ergs is the energy injected by a single supernovae, and $\nu = \nu_{100}/ (100 \text{M}_{\odot})$ is the number of supernovae per unit mass of star formation. We assume $\nu_{100} = 1$, meaning that one supernovae occurs for every $100 \text{M}_{\odot}$ of stars produced. For a Salpeter initial mass function (IMF), $\nu_{100} = 1.18$, and for a Chabrier IMF, $\nu_{100} = 1.74$ (\cite{1999ApJS..123....3L}, \cite{2009ApJ...697.2030S}). Then $\alpha = \alpha' \epsilon_{51} \nu_{100}$ is how we parametrize the energy injection, following \cite{2014ApJ...784...93Z} to most easily compare to their results using the CC model. In the CC model, this means $Q = \alpha' \dot{E}_{SN}/V$, whereas in our model, this parameterization means the energy per mass injected to the wind is 
\begin{equation}
\eta = 5.032 \times 10^{15} \alpha \nu_{100}
\end{equation}
and $\tilde{\eta} = \eta/ (GM/R)$. 

The $\tilde{\eta} = 0$ case corresponds to a polytropic wind, i.e. a wind obeying a law of the form $T \propto \rho^{\gamma-1}$, $P \propto \rho^{\gamma}$. For this scenario, supernovae heat the gas to a certain central temperature, $T_{0}$; however, after this central heating occurs, the wind is then driven solely by adiabatic expansion and advection, not from any additional energy deposition into the wind following a $1-r^{2}/R^{2}$ profile, as would be the case for $\tilde{\eta} > 0$ in our model.

\subsection{Determining the Critical Points}
\label{determineCritPoints}
To get the wind properties (velocity, temperature, density, etc.) as a function of radius, one must simultaneously solve the mass continuity, momentum, and energy equations, i.e. eqns. (\ref{massEQN}), (\ref{momEQN}), and (\ref{myTEQN}). Due to the nonlinearity of the equations, solving for the steady-state solution is best done numerically by integrating inwards to $r \approx 0$ from a known critical point, where the critical velocity, temperature, etc. are known, and also outwards to greater radii. 

Finding the critical point for a polytropic wind is straightforward. When adding additional energy ($\eta > 0$), it is not as straightforward to find the critical point. We follow the method outlined in \cite{1999isw..book.....L}: One writes the momentum equation as $F(\tilde{r},\tilde{u}, \tilde{T}, d\tilde{u}/d\tilde{r}, d\tilde{T}/d\tilde{r}) = 0$ and solves the following set of equations

\begin{equation}
\label{FEQN}
F(\tilde{r}, \tilde{u}, \tilde{T}, d\tilde{u}/d\tilde{r}, d\tilde{T}/d\tilde{r}) = 0
\end{equation}
\begin{equation}
\label{GEQN}
G \equiv  \frac{\partial F}{\partial \tilde{r}} + \frac{\partial F}{\partial \tilde{u}}\frac{d\tilde{u}}{d\tilde{r}} + \frac{\partial F}{\partial \tilde{T}}\frac{d\tilde{T}}{d\tilde{r}} = 0
\end{equation}
where du/dr can be found using L'Hospital's rule at the critical point. Because $F = 0$ all along the solution curve by definition, 
\begin{equation}
dF/d\tilde{r} =  G + \frac{\partial F}{\partial \tilde{u}'} \tilde{u}'' = 0
\end{equation}
all along the solution curve, as well. One can solve for the second derivative of velocity:
\begin{equation}
\tilde{u}'' = - \frac{G(\tilde{r},\tilde{u},\tilde{T})}{\frac{\partial F}{\partial \tilde{u}'}}
\end{equation} 
At the critical point, 
\begin{equation}
\frac{\partial F}{\partial \tilde{u}'} = 1 - \frac{\tilde{c_{s}}^{2}}{\tilde{u}^{2}} = 0
\end{equation} 
So for $\tilde{u}''$ to be well-defined at the critical point, $G(\tilde{r},\tilde{u},\tilde{T})$ should be exactly zero at the critical point. If this is true, then $d\tilde{u}/d\tilde{r}$ is continuous at the critical point, meaning that $\tilde{u}(r)$ will be a smooth function at the critical point. We would like this to occur, so we search for critical points by solving $F = 0$ and $G = 0$ simultaneously. 

\begin{figure}[!hb]
\label{manyTransonicSolns}
\centering
\hspace*{-0.6 cm}\includegraphics[width=0.57\textwidth]{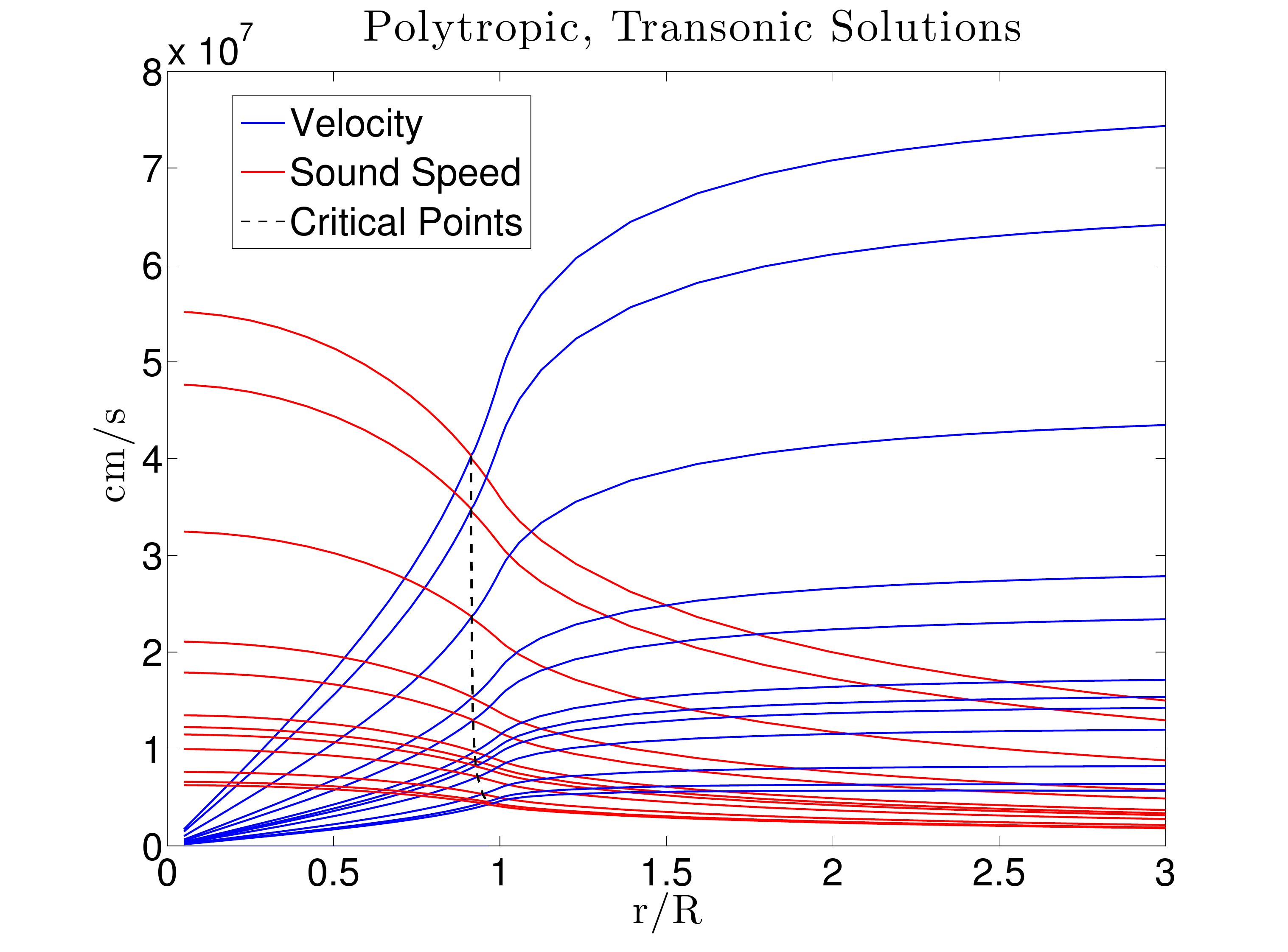}
\caption{$M = 10^{9} \text{M}_{\odot}$, $R = 200$ pc. Many polytropic ($\gamma = 5/3$), transonic wind solutions, each corresponding to a different central sound speed and, hence, different central temperature. Central wind temperatures range from $1.1 \times 10^{5} - 1.3 \times 10^{7}$ K.   Each wind is driven purely thermally, with no additional energy or mass injection. Wind velocities monotonically trend higher for higher central wind temperatures; therefore, solutions with high critical and asymptotic velocities can only be achieved with a high central temperature.}
\end{figure}

\begin{figure}[!ht]
\label{uc_vs_rc_varyKsi}
\centering
\includegraphics[angle = 0,width=0.5\textwidth]{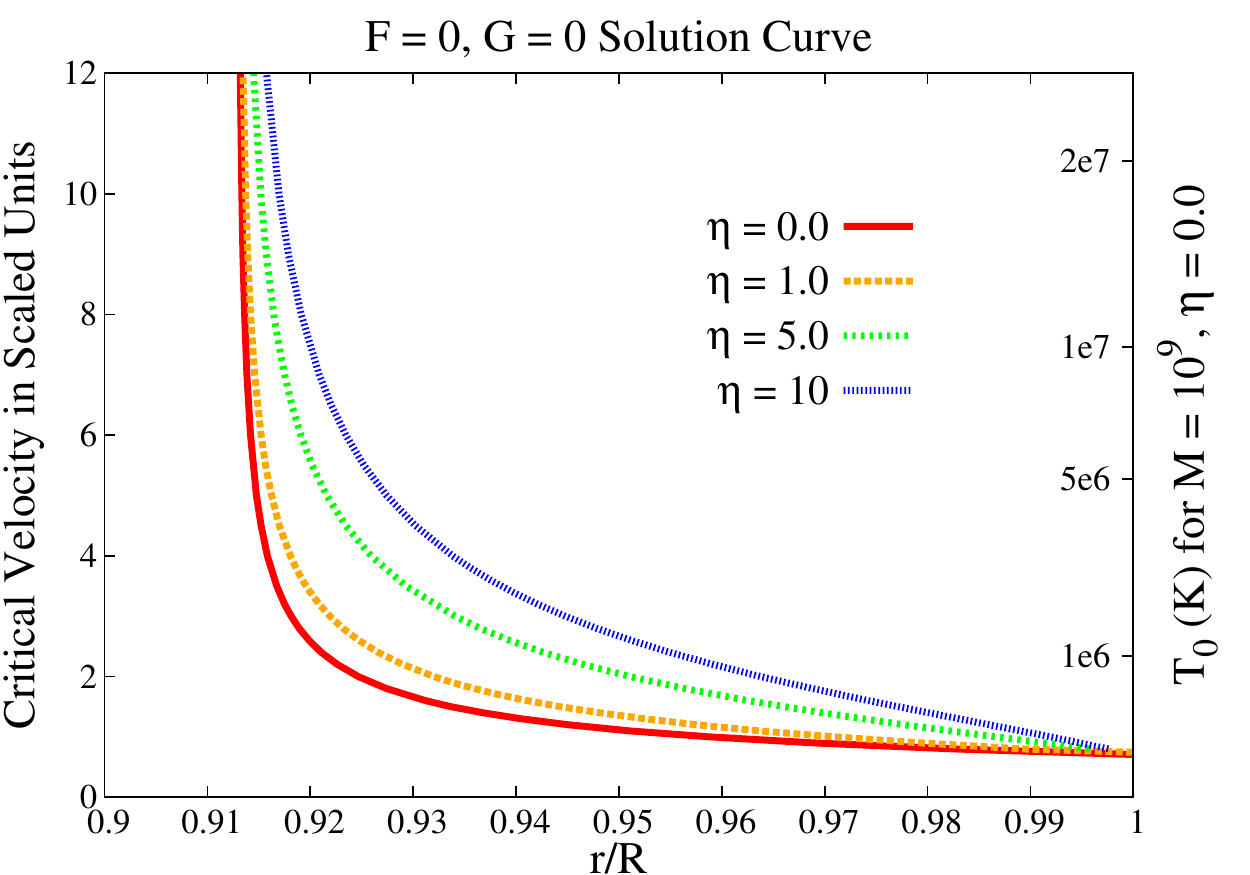}
\caption{Critical point solution sets for varying energy per mass inputs $\eta$ ranging from 0 to 10. The critical points are found by solving the equations $F = 0$ and $G = 0$. These two curves are the same; therefore, there are many critical points with smooth velocity profiles, each for a different central wind temperature. To get unscaled velocity, multiply scaled velocity by $\sqrt{GM/R}$. The right axis shows the corresponding central wind temperature for a galaxy of $10^{9} \text{M}_{\odot}$ and $\eta = 0.0$}
\end{figure}

Solving eqns. (\ref{FEQN}) and (\ref{GEQN}) without specifying the initial conditions for the desired transonic solution yields the same overlapping solution curves, i.e. an infinite number of smooth critical points, each corresponding to different initial conditions, as seen in Fig. \ref{manyTransonicSolns}. Fig. \ref{uc_vs_rc_varyKsi} gives the set of critical points for varying values of $\tilde{\eta}$, showing that the critical velocity generally shifts upwards with an increase in energy addition. It should be noted that the critical point locations for non-radiative winds do not depend on our mass injection parameter $q_{0}$; however, the $1-\frac{r^{2}}{R^{2}}$ profile is important. 

\subsection{Effect of Radiative Losses}
\label{radiative}

\begin{figure*}[!!ht]
\begin{minipage}{.5\textwidth}
\label{TwoSolsforLx}
\centering
\includegraphics[scale = 0.68, angle = 0]{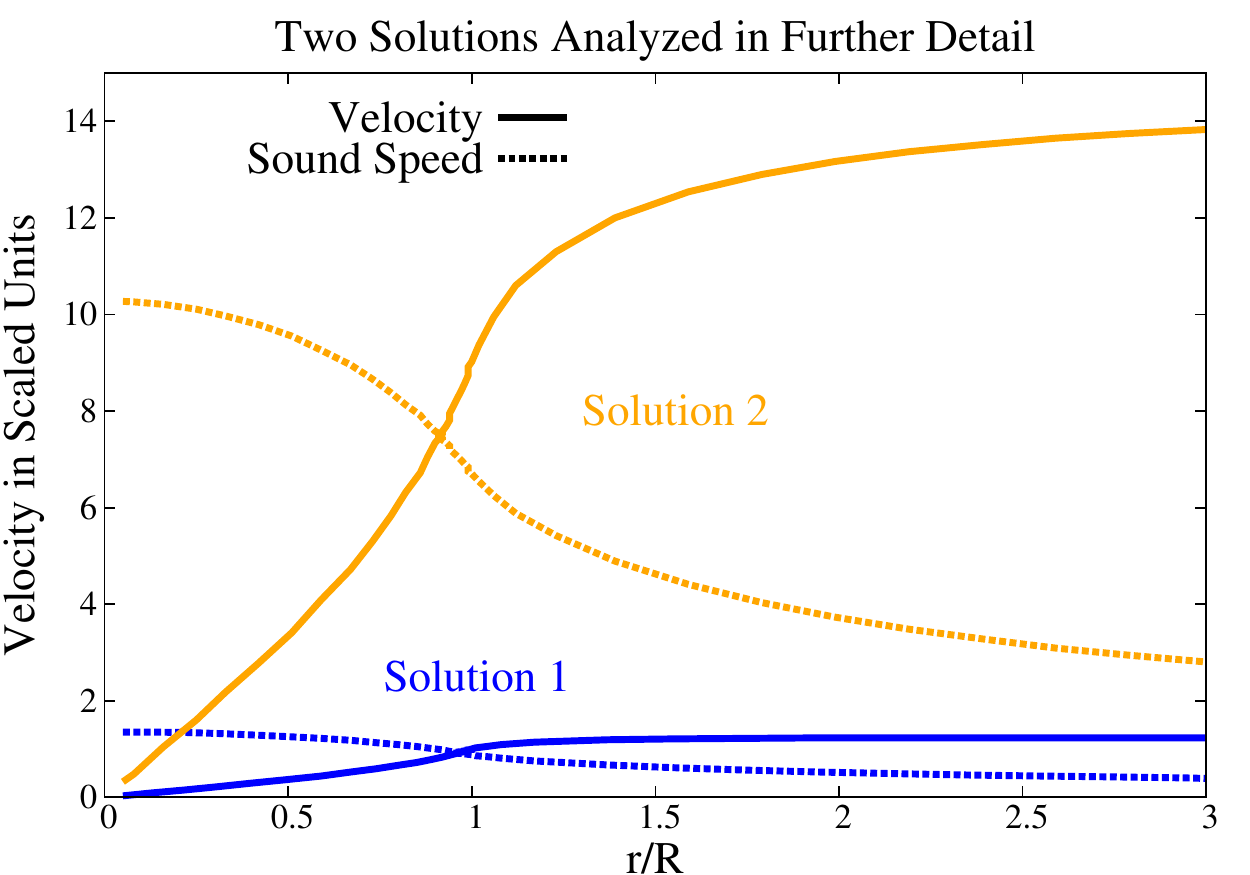}
\end{minipage}
\begin{minipage}{.5\textwidth}
\label{tcool}
\centering
\includegraphics[scale = 0.68, angle= 0]{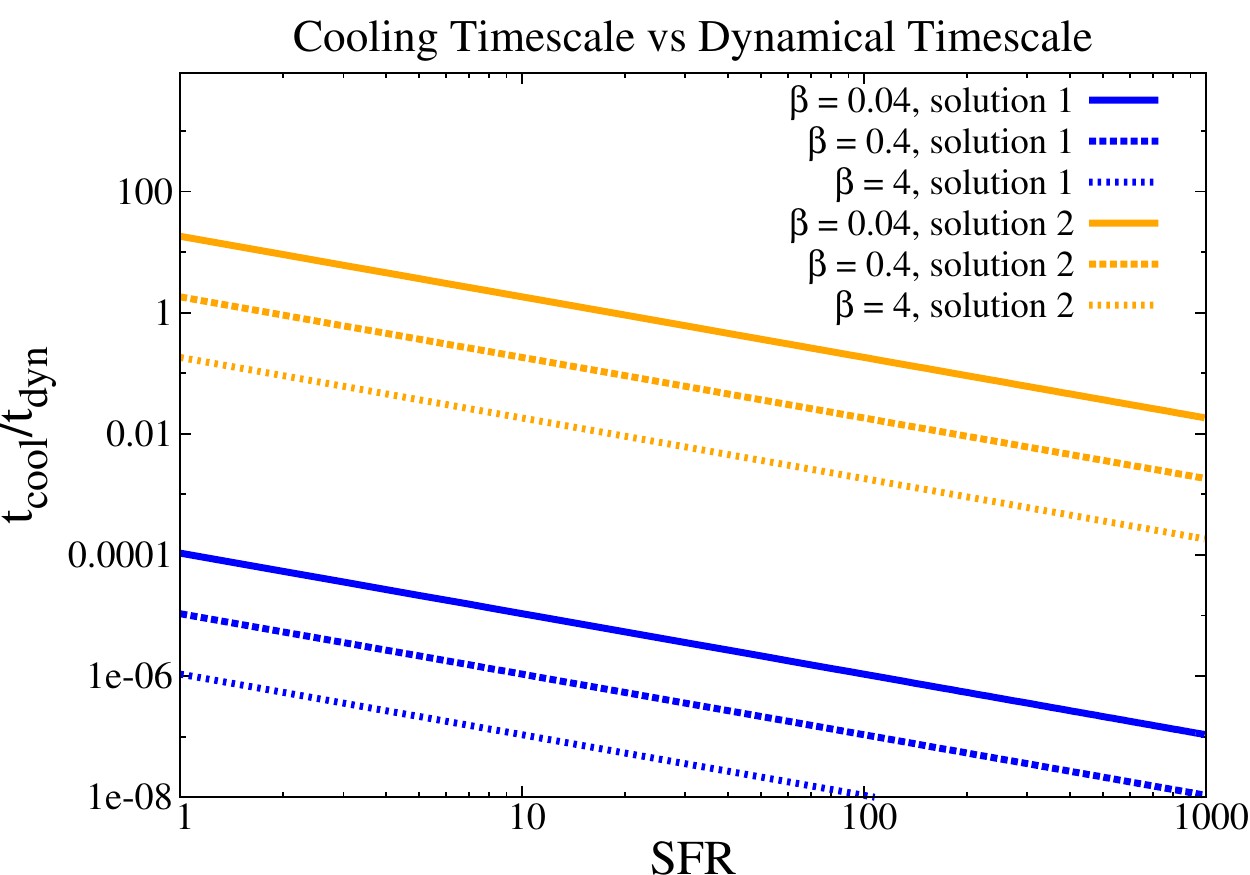}
\end{minipage}
\caption{Two transonic solutions (\textit{left panel}) and the ratio of the cooling timescale to the dynamical timescale of each solution over a range of SFRs for various $\beta$ and $M = 10^{9} \text{M}_{\odot}$ (\textit{right panel}). The lower velocity and lower sound speed solution is hereafter called solution 1, and the higher velocity and sound speed solution is hereafter called solution 2. To get unscaled velocity, multiply scaled velocity by $\sqrt{GM/R}$. The cooling timescale for winds in our model trends lower for higher SFR and also for higher $\beta$, both of which increase the density of the wind; however, we also see that radiative cooling is not nearly as important if we use a scaled solution with higher initial temperature (solution 2) than if we use one with lower initial temperature (solution 1).}
\end{figure*}

We further examine two polytropic, non-radiative wind solutions (shown in Fig. \ref{TwoSolsforLx}), and we show the ratio of the cooling timescale, $t_{\text{cool}} = 3k_{b}T/n\Lambda$, to the dynamical timescale, $t_{\text{dyn}} = \int_{0}^{R} dr/u$, of the solutions for various star formation rates and mass-loading efficiencies. This provides a first order estimate as to whether radiative losses should be included. The cooling timescale trends lower for lower SFR and also for higher $\beta$; however, we also see that radiative cooling is not nearly as important if we use a scaled solution with higher initial temperature (solution 2) than if we use one with lower initial temperature (solution 1). Nevertheless, it is important to include radiative losses for all temperature ranges. 

With radiative losses included, our mass continuity equation stays the same, but our momentum and temperature equations now contain radiative loss terms: 

\begin{equation}
\label{radiative_momentumEqn}
    \rho u \frac{\partial u}{\partial r} = -qu - c_{s}^{2} \frac{\partial \rho}{\partial r} - \rho \frac{\partial \Phi}{\partial r} + (\gamma-1) \frac{\rho^{2} \Lambda(T)}{\overline{m}^{2} u}
\end{equation}

\begin{multline}
\label{radiative_tempEqn}
u\frac{k_B}{\overline{m}}\frac{dT}{dr}= (\gamma - 1)\frac{dQ}{dt} \\
 +(\gamma - 1)\frac{k_BTu}{\overline{m}\rho}\frac{d\rho}{dr} - (\gamma-1) \frac{\rho \Lambda(T)}{\overline{m}^{2}}
\end{multline}
where $\rho/\overline{m} = n$ is the number density, $\overline{m} = 1.021 \times 10^{-24}$ g is the mean mass of a wind particle, and $\Lambda(T)$ is the optically thin radiative loss function.

\begin{figure}[!ht]
\label{analyticCooling}
\centering
\includegraphics[scale = 0.34, angle= -90]{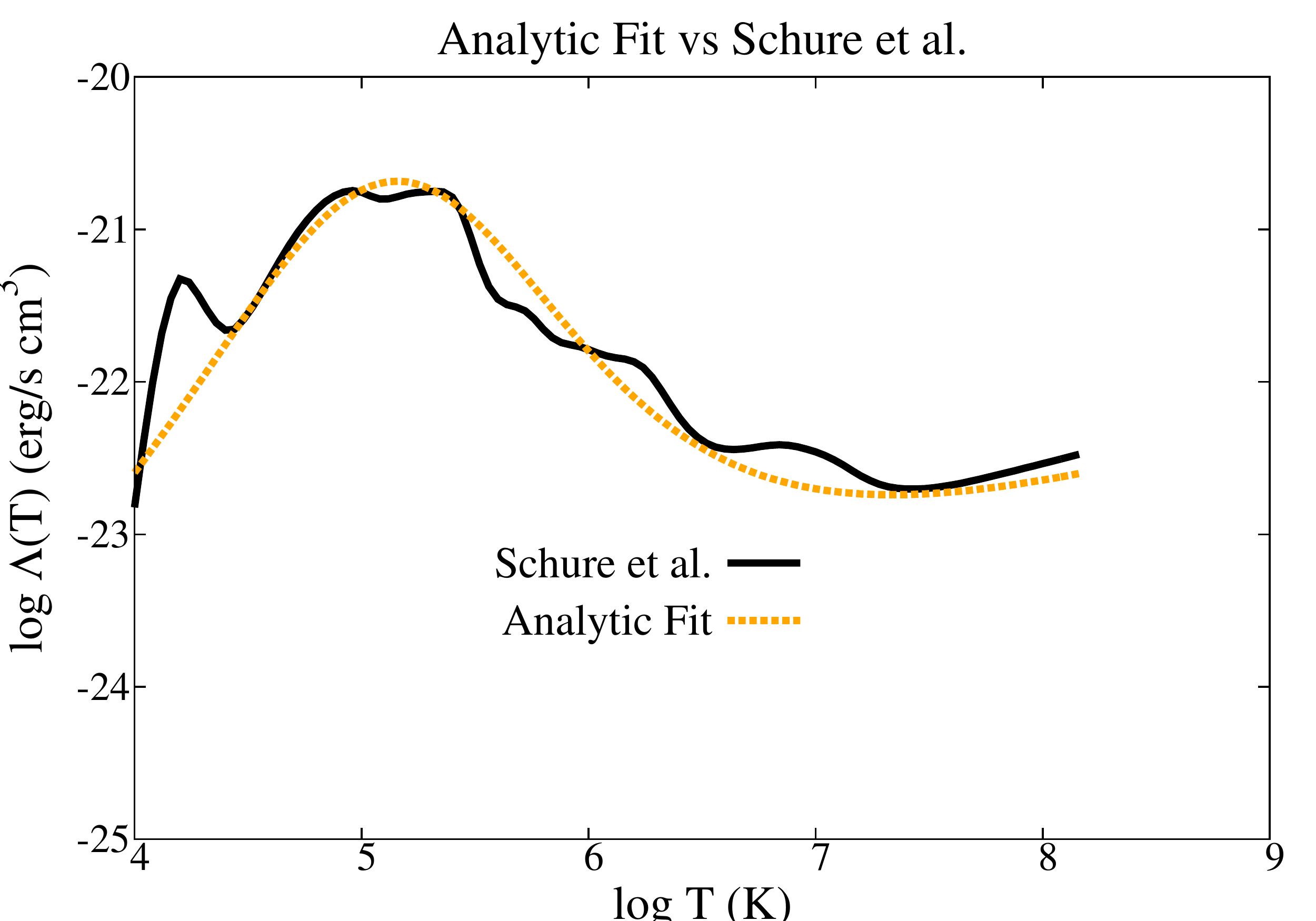}
\caption{To self-consistently include radiative losses in our model, we choose to use the analytic fit from \cite{2012ApJ...755...93I}, which fairly well fits the peak near $5 \times 10^{5}$ K as well as the structure at higher temperatures. It also captures the steep drop in emissivity for temperatures below $10^{5}$ K; however, our assumption of CIE should not be trusted below about 5000 K. For low temperatures, one would need to include a non-equilibrium model.}
\end{figure}

Instead of using a more detailed cooling curve such as from \cite{2009A&A...508..751S}, we choose for simplicity to use an analytic fit very similar to that of \cite{2012ApJ...755...93I}, given by eqn. (\ref{analyticCoolingFCN}), which fits the peak near $5 \times 10^{5}$ K fairly well and also captures the behavior for both higher and lower temperatures, assuming collisional ionization equilibrium (CIE)
\begin{equation}
\label{analyticCoolingFCN}
\Lambda(T) = \Lambda_{0}10^{\Theta(\theta)}
\end{equation}
where $\theta(T) = \text{log}_{10}(T/ T_{0})$, $T_{0} = 2 \times 10^{5}$ K, and we choose $\Lambda_{0} = 1.5 \times 10^{-21}$ ergs $s^{-1} cm^{3}$
\begin{multline}
\Theta(\theta) = 0.4\theta - 3 \\ + 6.2/[\text{exp}(1.5\theta + 0.08) + \text{exp}(-(\theta + 0.08))]
\end{multline}
A plot of the analytic solution vs more detailed calculations from \cite{2009A&A...508..751S} is given in Fig. \ref{analyticCooling}.

\begin{figure}[!h]
\label{radiative_critPoints}
\centering
\includegraphics[scale = 0.35, angle = -90]{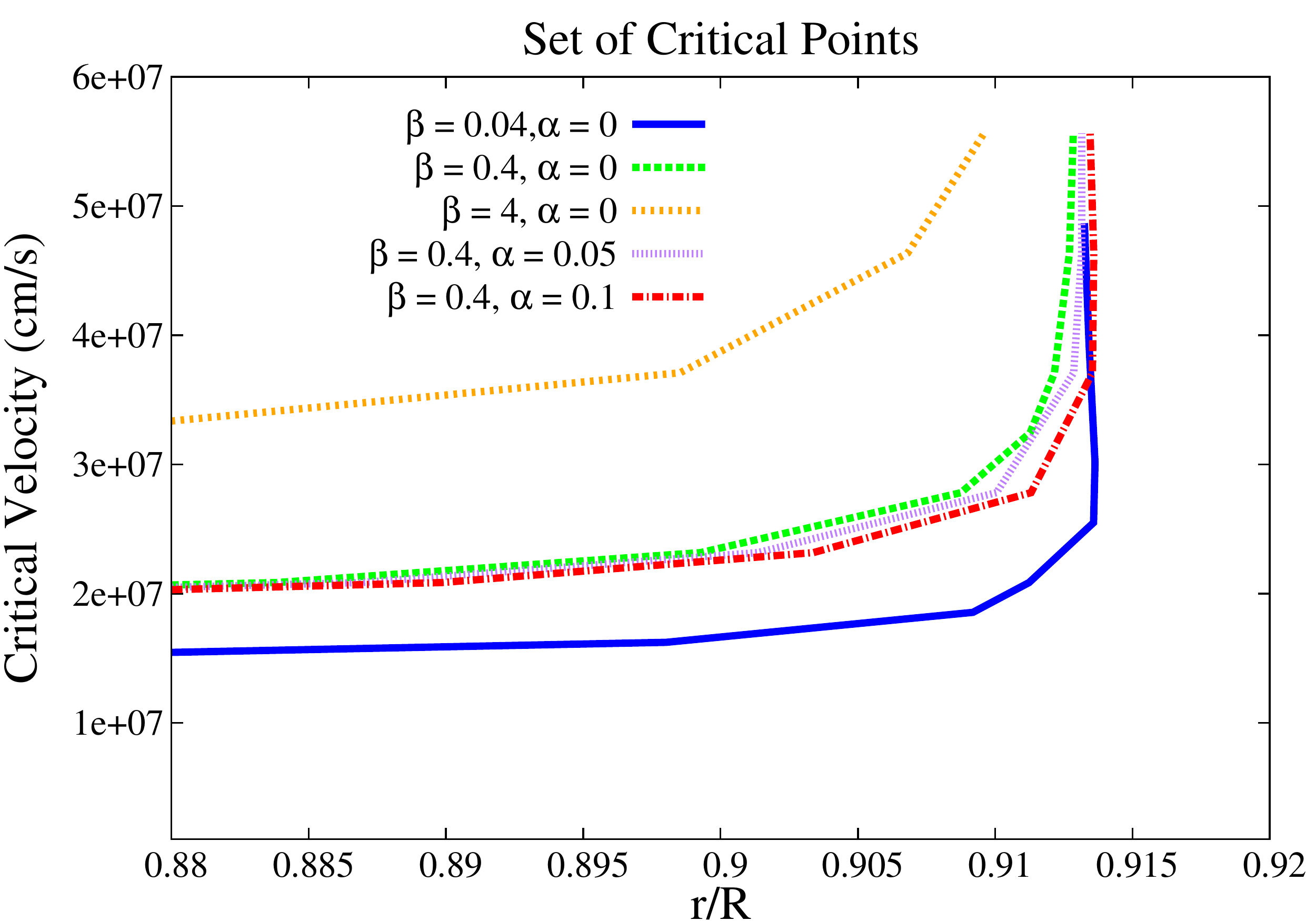}
\caption{$\text{SFR} = 1 \text{M}_{\odot}/yr$, galaxy mass $M = 10^{9} \text{M}_{\odot}$. These are solution curves for critical points of radiative winds. One can observe that higher critical velocities, and hence higher initial temperatures, are required to achieve a transonic solution with higher $\beta$.}
\end{figure}

\begin{figure}[!ht]
\label{centralTemp}
\centering
\hspace*{-0.9 cm}\includegraphics[scale = 0.41]{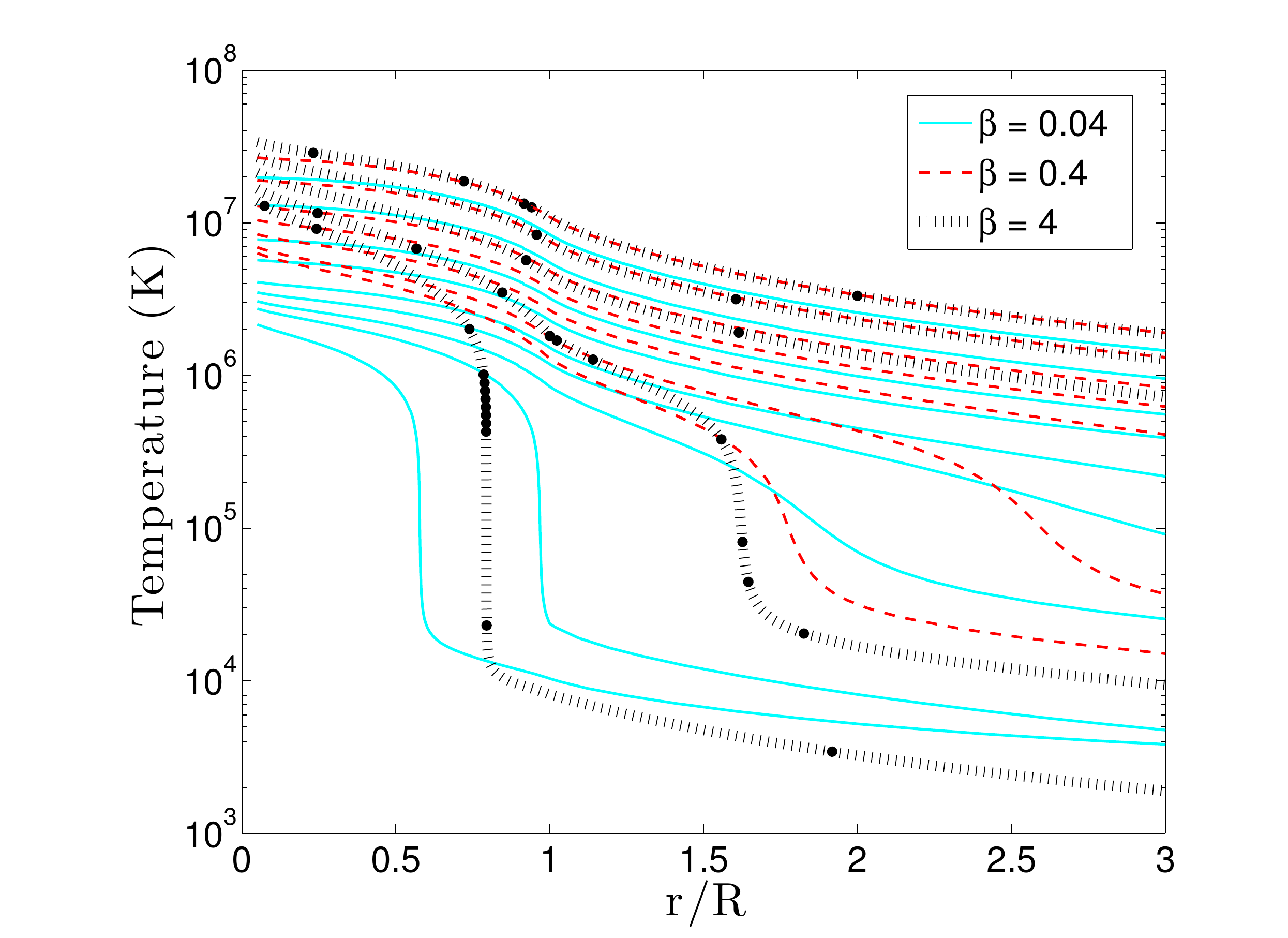}
\caption{$SFR = 1 \text{M}_{\odot}\text{/yr}$, galaxy mass $M = 10^{9} \text{M}_{\odot}$. Wind temperature profiles corresponding to a representative sample of critical points in Fig. \ref{radiative_critPoints} with no additional energy input  ($\alpha = 0$). In general, the low-$\beta$ transonic solutions can be driven at lower central temperatures, whereas the high-$\beta$ solutions can only be driven at higher temperatures. The existence of low temperature winds for which the wind temperature plummets to close to $10^{3}$ K will be discussed in Section \ref{lowTemp}.}
\end{figure}

For now, we just consider transonic solutions for radiative winds. The system of steady-state, radiative wind equations is given by eqns. (\ref{radiative_momentumEqn}) and (\ref{radiative_tempEqn}). As for the non-radiative case, we scale these equations by $GM/R$ and solve for the critical points, i.e. we solve $F(\tilde{r},\tilde{u}, \tilde{T}, d\tilde{u}/d\tilde{r}, d\tilde{T}/d\tilde{r}) = 0$ and
$G \equiv  \frac{\partial F}{\partial \tilde{r}} + \frac{\partial F}{\partial \tilde{u}}\frac{d\tilde{u}}{d\tilde{r}} + \frac{\partial F}{\partial \tilde{T}}\frac{d\tilde{T}}{d\tilde{r}} = 0$. As for the non-radiative case, we find that $F = 0$ and $G = 0$ define the same solution curves, which are given in Fig. \ref{radiative_critPoints} for various mass-loading factors $\beta$. Looking at Fig. \ref{radiative_critPoints}, we also see that the critical velocities tend greater with increasing $\beta$; an increased critical velocity is only achieved with an increased initial temperature, assuming the initial velocity always approaches $u = 0$ at $r = 0$. Therefore, we see that more heavily mass-loaded winds (higher $\beta$) require a higher initial temperature to achieve any transonic solution. This is more explicitly shown in Fig. \ref{centralTemp}.

\begin{figure}[!ht]
\label{radiative_varyBeta}
\centering
\includegraphics[scale = 0.35,angle=-90]{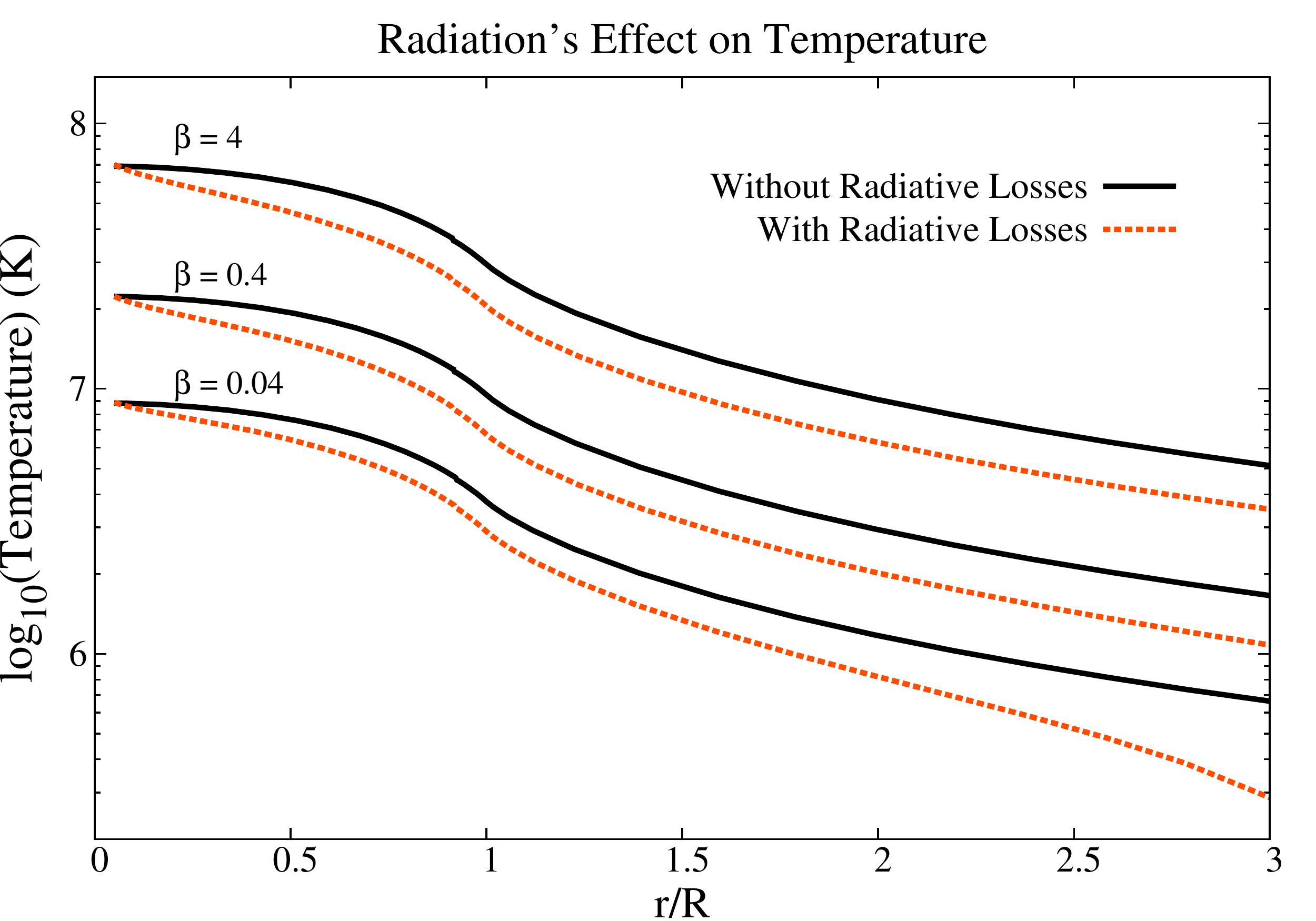}
\caption{Plot of temperature for $\beta = 0.04, 0.4, 4$. As shown in Fig. \ref{radiative_critPoints}, a more mass-loaded (higher $\beta$) wind requires a higher central temperature to achieve a transonic solution. Comparing radiative to non-radiative winds, we see that the effect of radiation is more significant in higher mass-loaded winds (due to higher density in the wind) than the lower mass-loaded winds; however, the difference is not extreme because the high-$\beta$ winds are also those with high temperature. Conversely, the low-$\beta$ solution has lower density; however, it also has a lower temperature, meaning that radiative effects can still be important even if the wind is not very dense.}
\end{figure}

Recall from Fig. \ref{tcool} that greater mass-loading for the same initial temperature (same solution curve) implies a more radiative wind due to an increase in density; however, because the cooling time is proportional to the temperature, higher temperature winds are generally less radiative. Therefore, since high-$\beta$, radiative winds can only become transonic if given a high temperature, the radiative losses incurred from heavily mass-loading the winds are somewhat lessened by the higher temperature. Conversely, since it is possible to have low-temperature, low-$\beta$ transonic solutions, radiative effects can also be important in low mass-loaded winds. 

We show this in Fig. \ref{radiative_varyBeta} by comparing two transonic winds with the same initial temperature: one with radiative loss physics included and one without for various mass-loading factors. The resulting velocity is decreased by about a factor of two at large radii, the temperature decreases by a factor of $2-4$, and the density is slightly increased for large radii after starting off with a lower central density for the radiative solutions. A plot of temperature is given in Fig. \ref{radiative_varyBeta}. 

\section{Efficiency of Winds in Expelling Mass}
\label{efficiency}
We have modified the widely-used Chevalier and Clegg wind model by including an extended mass source term, a simplified gravitational potential, and radiative losses using an analytic cooling curve that fairly well encapsulates the behavior of more detailed cooling curves at all temperatures. With these logical modifications, we believe our model includes all the necessary physics while still being quite simple, allowing us to probe the fundamental questions of importance to observers and the galaxy formation community: What conditions are necessary for a thermally driven wind? How efficiently can mass be expelled from a galaxy by a thermally driven galactic wind? Are subsonic winds or transonic winds more efficient outflows? 

The energy per unit time driven outward by a wind is given by 
\begin{equation}
\frac{f}{4\pi} = \dot{M}\left[\frac{1}{2}u^{2} + \frac{\gamma}{\gamma-1}\frac{P}{\rho} \right]
\end{equation} 
excluding gravitational potential and radiative loss terms that decline with distance and do not contribute asymptotically. 

Using $\dot{M} = \beta \text{SFR}$,
\begin{equation}
\frac{f}{4\pi} = \beta \text{SFR}\left[\frac{1}{2}u^{2} + \frac{c_{s}^{2}}{\gamma-1} \right]
\end{equation} 

We define the energy per mass quantity 
\begin{equation}
\epsilon \equiv \frac{1}{2}u^{2} + \frac{c_{s}^{2}}{\gamma-1}
\end{equation}

This effectively quantifies how efficiently the winds can expel mass and energy from the galaxy.

\begin{figure*}[!!ht]
\label{solutionsEdot}
\centering
\begin{minipage}{.4\textwidth}
\hspace*{-2.4cm}\includegraphics[scale = 0.32]{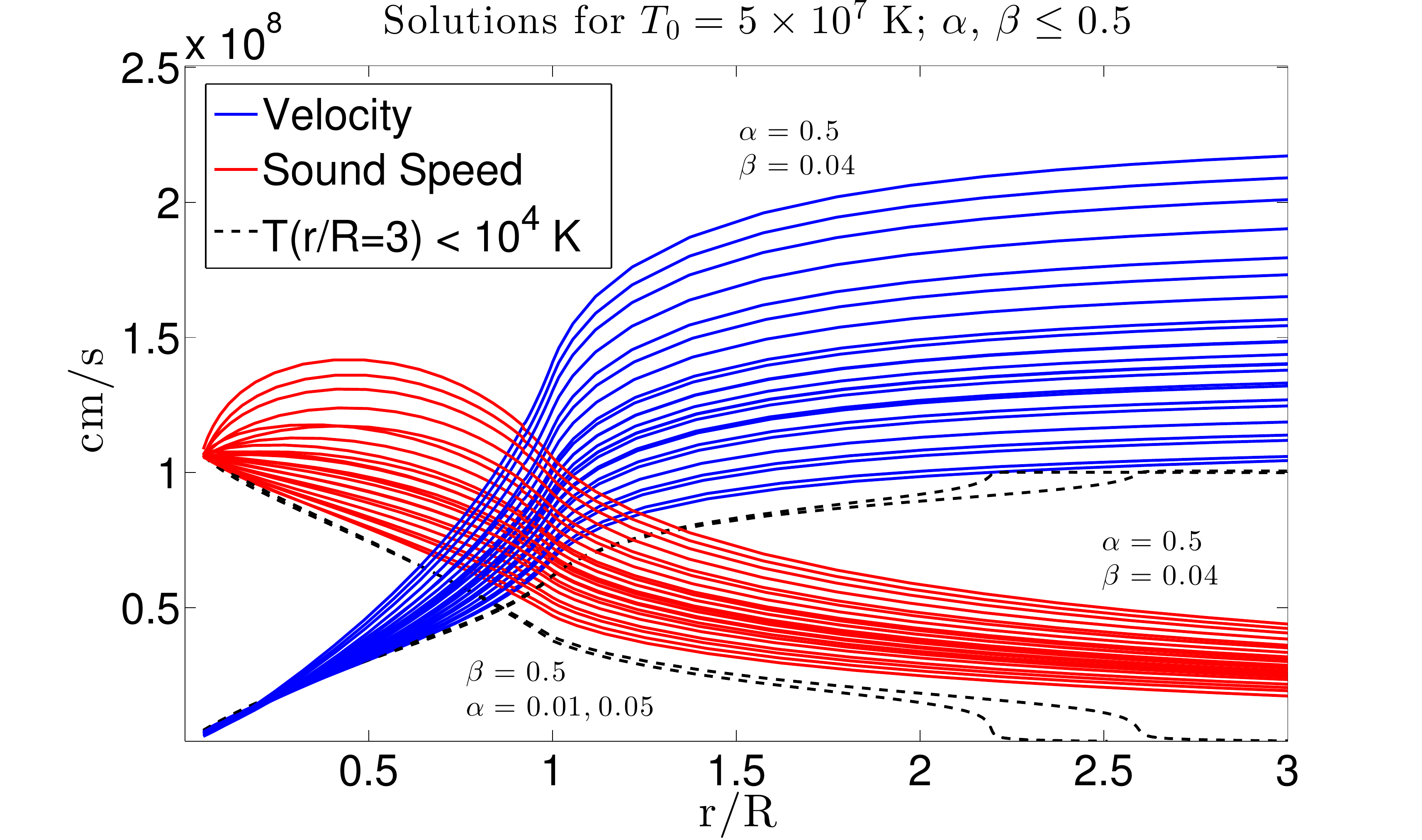}
\end{minipage}
\begin{minipage}{.4\textwidth}
\hspace*{0.5cm}\includegraphics[scale = 0.30]{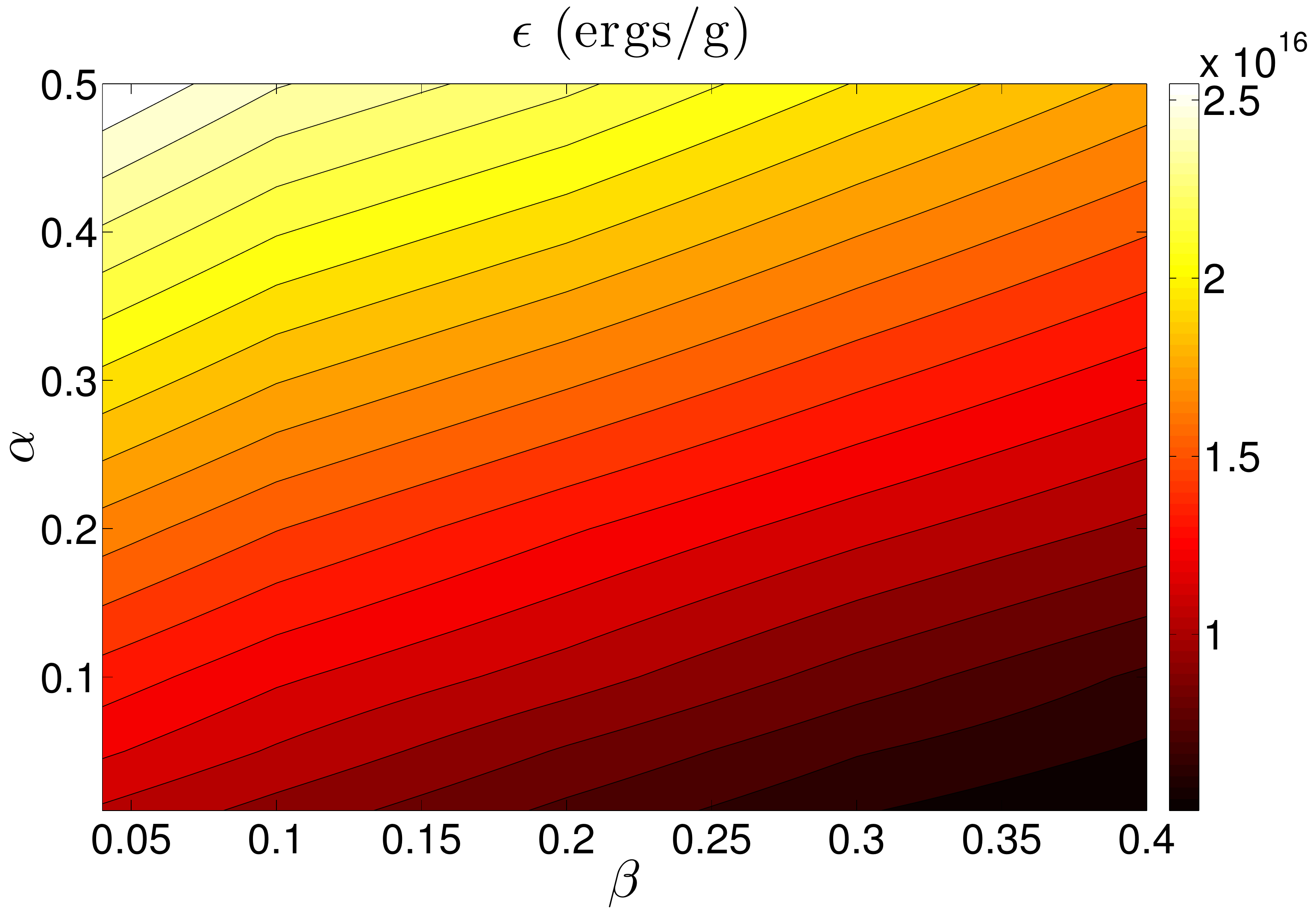}
\end{minipage}
\caption{Plot of various radiative wind solutions (\textit{left panel}) and their efficiencies (\textit{right panel}) according to our measure, $\epsilon$, of energy per mass driven outward by the wind. Each wind has a central temperature of $5 \times 10^{7}$ K, a SFR of $100 \text{M}_{\odot}\text{/yr}$, a galaxy mass of $10^{9} \text{M}_{\odot}$, and a galaxy radius of $R = 200$ pc. $\beta$ ranges from 0.04 to 4, and $\alpha$ ranges from 0 to 0.5. Winds with the highest velocity (blue curve) also have the highest sound speed (red curves). These high-velocity, high-temperature winds are the most efficient, and correspond to the upper left corner of the efficiency plot in the right panel. At high $\beta$, radiative losses are more important due to the wind being denser, resulting in a lower temperature and lower velocity wind. Therefore, $\epsilon$ decreases as $\beta$ increases. This can be offset by a large $\alpha$, however, as $\epsilon$ generally increases as $\alpha$ increases. For winds with $\beta \agt 0.4$ and $\alpha \alt 0.5$, the temperature drops below $10^{4}$ K within $600$ pc of the galactic center, making our assumption of CIE invalid and any calculation of $\epsilon$ meaningless. A few of these solutions are plotted with dashed lines in the left panel.}
\end{figure*}

\subsection{Efficiency of Transonic Outflows}
\label{efficiency1}

We have shown that for each mass-loading factor $\beta$ and energy factor $\alpha$, there are infinitely many transonic solutions, each for a different central wind temperature, which can drive mass to large radii. The question we would like to ask is ``for the same central wind temperature, what are the trends in $\alpha$ and $\beta$ that lead to the most efficient mass expulsion from a galaxy?" We look at this by considering a set central temperature of $5 \times 10^{7}$ K, a SFR of $100 \text{M}_{\odot}\text{/yr}$, and a galaxy mass of $10^{9} \text{M}_{\odot}$, which would be fairly typical for an active starburst galaxy, and we vary $\alpha$ and $\beta$. We then calculate $\epsilon$ at a radius of $r/R = 3$, or $600$ pc, at which point the kinetic energy dominates the energy flux, for each radiative wind solution to track the energy per mass of the wind. The solutions considered and a contour plot of energy per mass for the $\beta \le 0.4$ are shown in Fig. \ref{solutionsEdot}.

At high $\beta$, radiative losses are more important due to the wind being denser, resulting in a lower temperature and lower velocity wind. Therefore, $\epsilon$ decreases as $\beta$ increases. This can be offset by a large $\alpha$, however, as $\epsilon$ generally increases as $\alpha$ increases. We also see that, unless the wind has a high $\alpha \agt 0.5$, solutions with $\beta \agt 0.4$ drop in temperature below $10^{4}$ K very sharply within $600$ pc, at which point our assumption of CIE becomes invalid, and our calculation of $\epsilon$ shouldn't be trusted. Including radiative heating at temperatures $\alt 10^{4}$ K will be left to further work.

Something to note about the sound speed profiles in Fig. \ref{solutionsEdot} is that, for high $\alpha$, the sound speed, and hence temperature, increases near the galactic center. This is because energy is being injected at a high rate, and advection and radiative losses cannot cool the gas fast enough. This is an interesting result because the $1-r^{2}/R^{2}$ profile for energy injection would seem logical, as you expect more supernovae to occur near the galactic center. We hope to include heat conduction in a future model, which should decrease this rise in temperature.

\subsection{Efficiency of Subsonic Outflows}
\label{subsonic}

\begin{figure*}[!ht]
\label{subsonicSolns}
\centering
\begin{minipage}{.4\textwidth}
\hspace*{-2.2cm}\includegraphics[scale = 0.34]{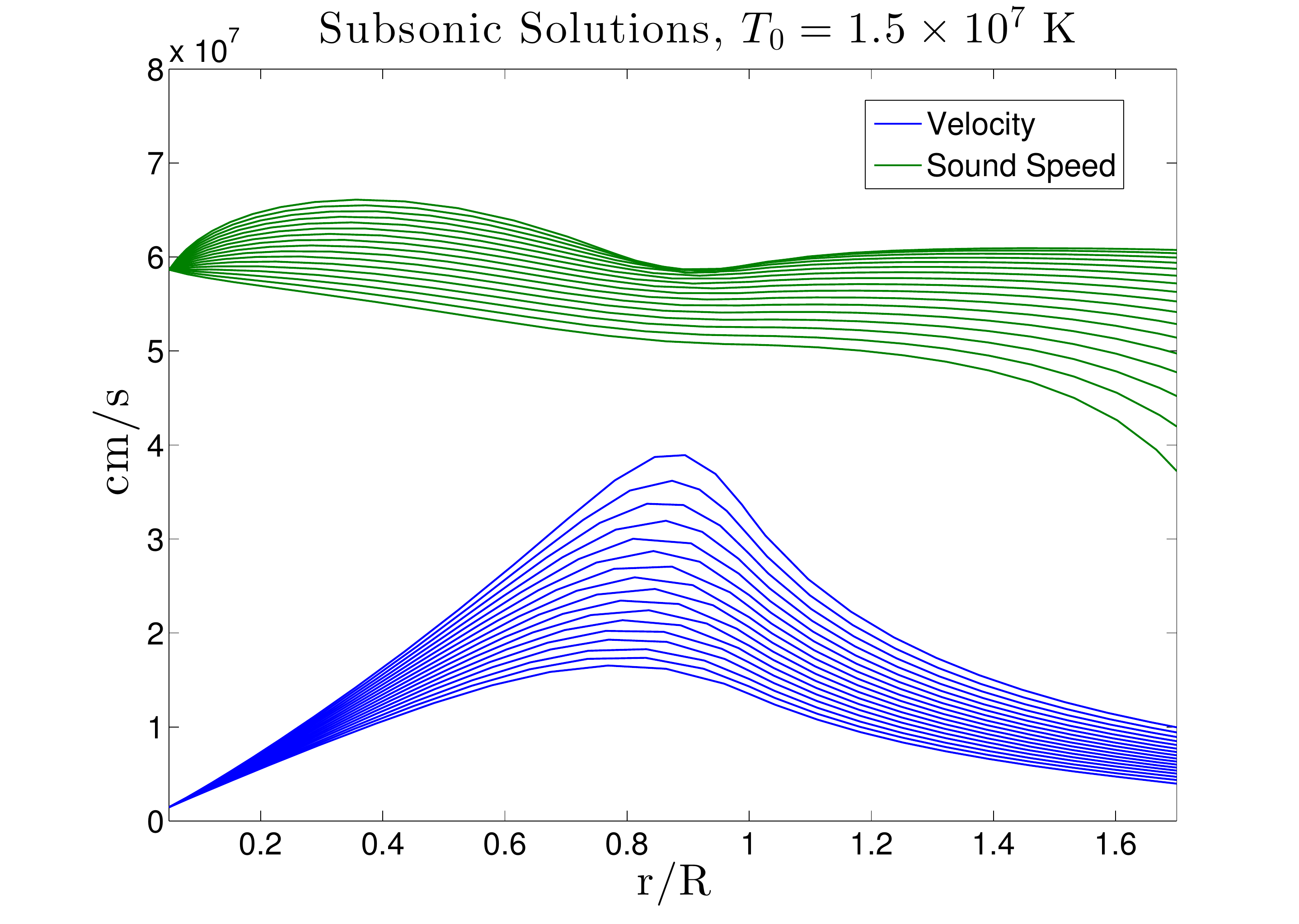}
\end{minipage}
\begin{minipage}{.4\textwidth}
\hspace*{-0.53 cm}\includegraphics[scale = 0.34]{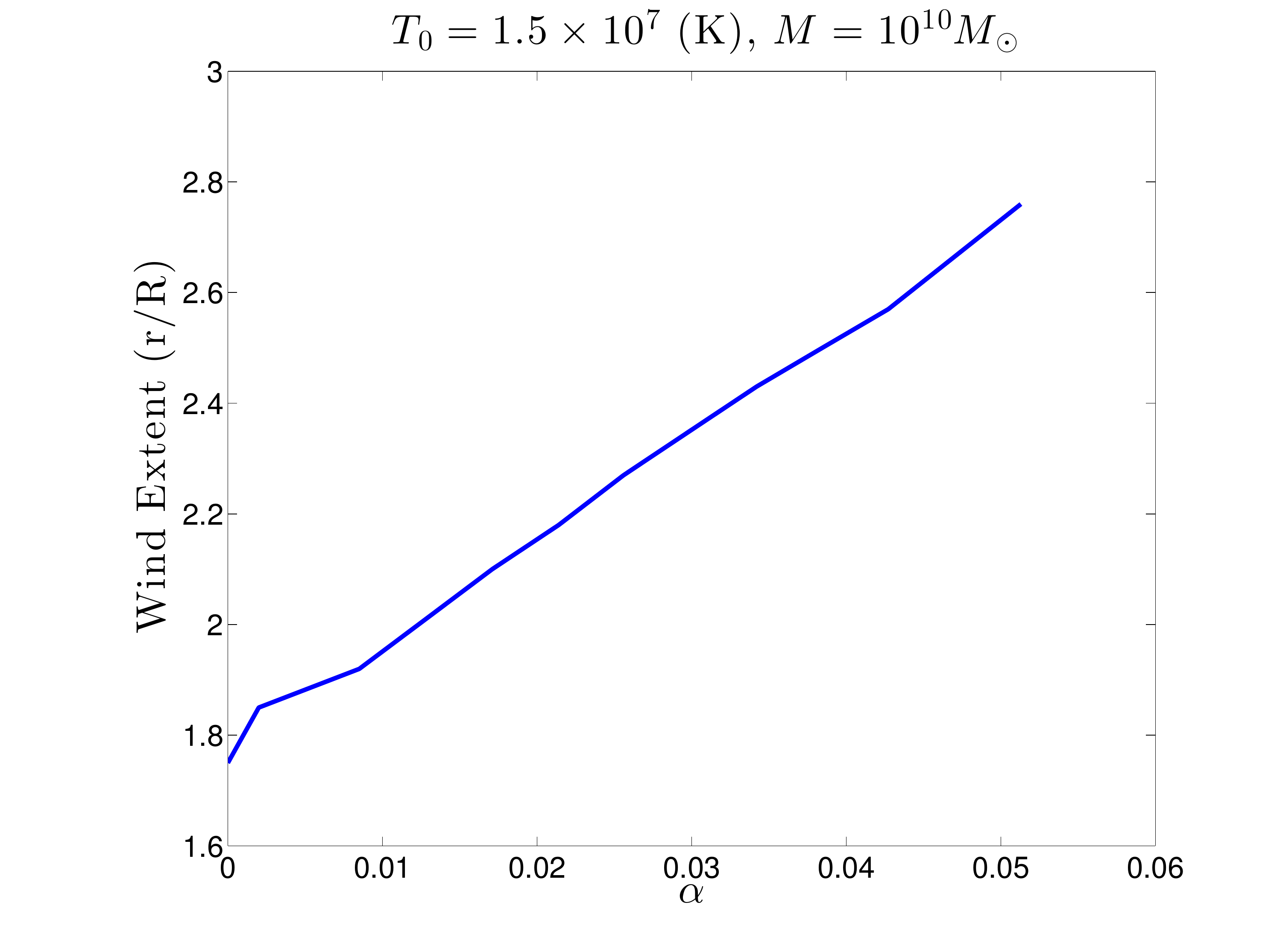}
\end{minipage}
\caption{$M = 10^{10} \text{M}_{\odot}$. Set of a few subsonic solutions with the same central temperature $T = 1.5 \times 10^{7}$ K, $\beta = 0.04$, but with different $\alpha$ (\textit{left panel}) ranging from 0 to 0.05. Also a plot of wind extent, i.e. how far these winds penetrate outside a galaxy until the temperature (and hence $\epsilon$) drops considerably (\textit{right panel}). The sound speed curves (green) and velocity curves (blue) in the left panel monotonically increase with increasing $\alpha$.}
\end{figure*}

We have so far only considered transonic solutions, but there are a wealth of subsonic solutions that could also expel mass from a galaxy, albeit at a slower pace since the asymptotic velocity is much lower. In Fig. \ref{subsonicSolns}, we show a set of subsonic solutions with various $\alpha$ and the same central temperature, $\beta = 0.04$, $\text{SFR} = 1 \text{M}_{\odot}\text{/yr}$, $M = 10^{10} \text{M}_{\odot}$. The lowest sound speed curve corresponds to the lowest velocity curve, for which $\alpha = 0$. For greater $\alpha$, the temperature and velocity curves are higher. Because radiative losses are greatest at low temperatures near the cooling curve peak, the low $\alpha$ (lower temperature) curves are most affected by radiative losses, causing a sharp drop in temperature at some radius. For the $\alpha = 0$ solution, this sharp drop occurs at roughly $r/R = 1.8$, meaning that $\frac{1}{2}u^{2} + \frac{c_{s}^{2}}{\gamma-1} = \epsilon$ drops by orders of magnitude at $\approx 360$ pc. In Fig. \ref{subsonicSolns}, we roughly track the point at which temperature and, hence, $\epsilon$ drops sharply for winds of various $\alpha$. We see that winds with higher $\alpha$ can expel mass to greater radii, but only so far. The transonic solution, however, is the only solution that can efficiently expel mass to very large radii. Therefore, we conclude that transonic solutions are indeed the most important, especially those with high $\alpha$ and low $\beta$.

\begin{figure*}[!ht]
\label{lowTempFig}
\centering
\begin{minipage}{.4\textwidth}
\hspace*{-1.8 cm}\includegraphics[scale = 0.32]{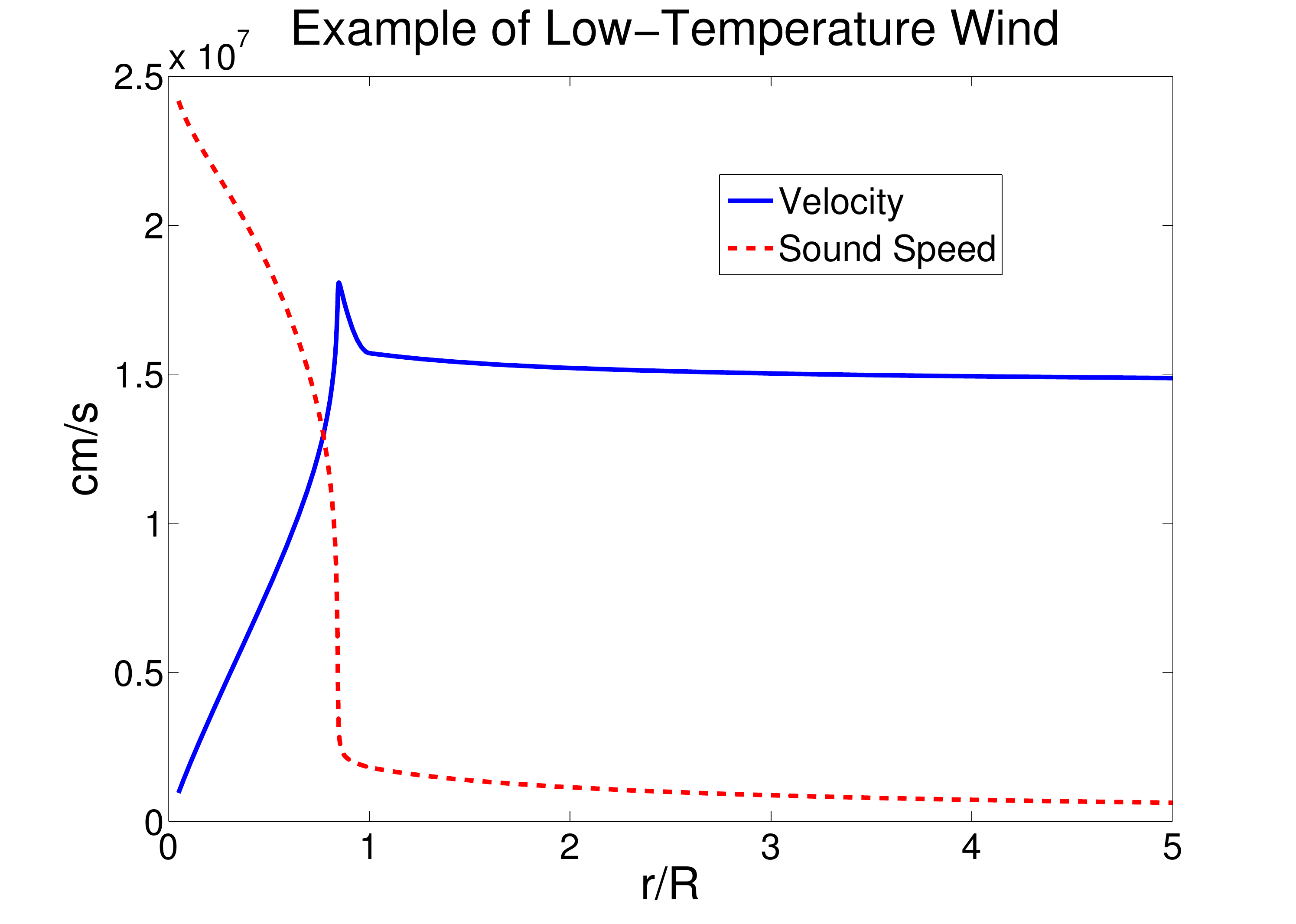}
\end{minipage}
\begin{minipage}{.4\textwidth}
\hspace*{-0.3cm}\includegraphics[scale = 0.30]{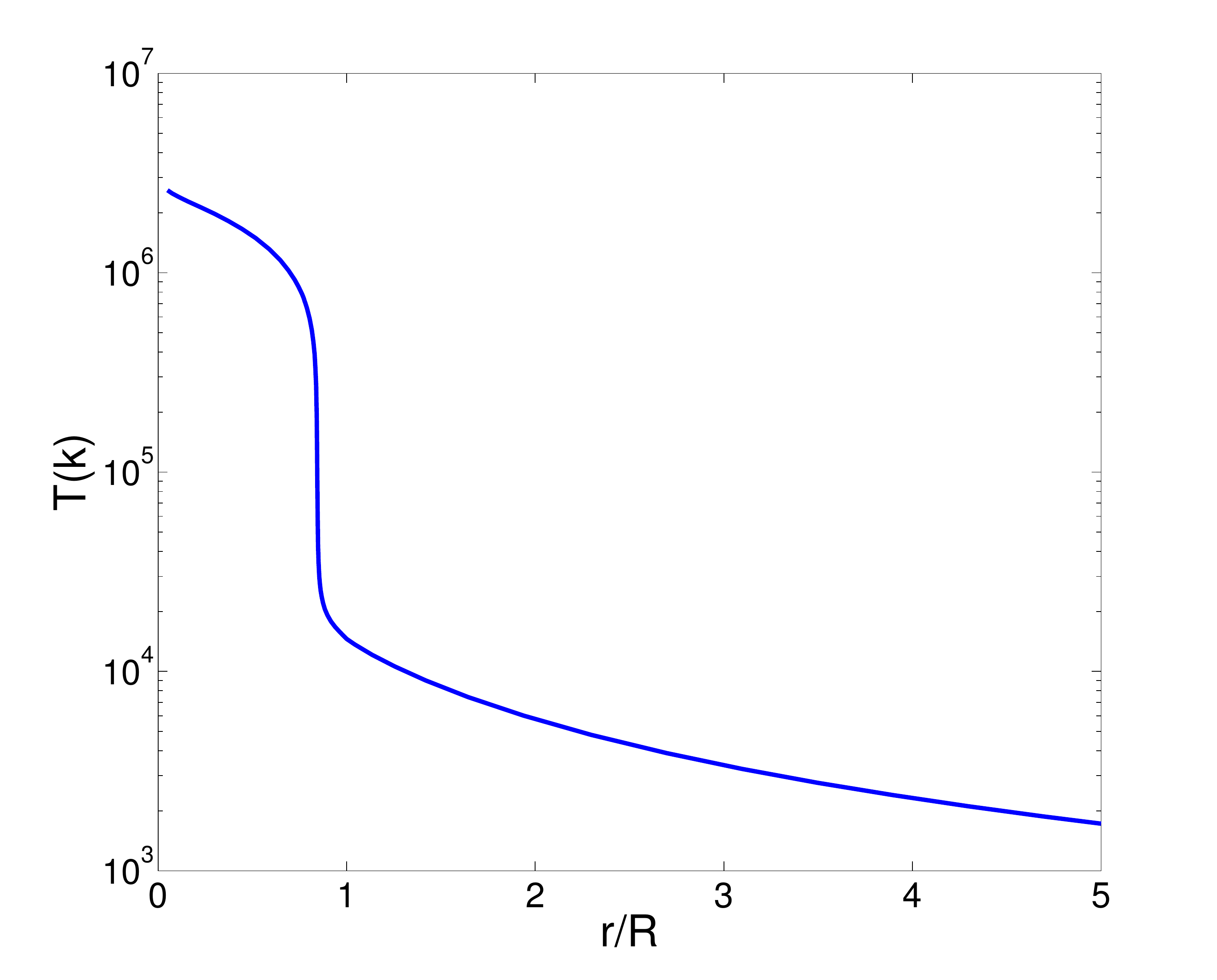}
\end{minipage}
\caption{An example of a low-temperature wind with $\beta = 0.04$, $\alpha = 0$ from an $M = 10^{9} \text{M}_{\odot}$, $\text{SFR} = 1 \text{M}_{\odot}\text{/yr}$ galaxy. Velocity and sound speed are plotted (\textit{left panel}) as well as temperature (\textit{right panel}). The wind temperature decreases very rapidly as the temperature nears the peak of the cooling curve, reaching the sonic point at just $r_{c}/R = 0.7725$, or $154.5$ pc assuming $R = 200$ pc. Once the temperature falls below the peak, the emissivity drops considerably very quickly, hence the quick flattening of the temperature profile and the quick deceleration of the wind. Radiative losses then become negligible, and the wind propagates adiabatically.}
\end{figure*}

\subsection{Low Temperature, High Velocity Outflows}
\label{lowTemp}

We notice from Fig. \ref{radiative_critPoints} that the sonic points for an outflow can be closer to the galactic center, but still near the edge, depending on the wind's central temperature. This is because low temperature winds cool more rapidly than high temperature winds further away from the cooling curve peak. Fig. \ref{lowTempFig} shows a transonic solution that cools very rapidly to temperatures in the range of $10^{4} - 10^{5}$ K, at which point the wind goes over the peak of the cooling curve, and the temperature profile flattens considerably. We have encountered many other such winds in our models that drop within a few hundred parsecs to even lower temperatures, at which point the assumption of CIE breaks down. These low temperature winds seem especially common in starburst galaxies with high SFR, such as in Fig. \ref{solutionsEdot}, where even winds of $\beta \approx 0.5$ can drop to temperatures $\alt 10^{4}$ K very quickly unless the wind is also highly energy-loaded with $\alpha \agt 0.5$. Despite our assumption of CIE breaking down at such low temperatures, we believe that radiative cooling in our model can explain the prevalence of cool, fast outflows from galaxies. This is in agreement with \cite{2011ApJ...743..120S}, as well as the recent work of \cite{2015arXiv150704362T} who see the same drastic radiative cooling for high-$\beta$ winds using a different modification of the CC model. For low $\beta$, our results also qualitatively agree in that the wind expands approximately adiabatically, as radiative cooling is not as important (see \cite{2015arXiv150704362T}, Fig. 2). They also find that photoionization equilibrium (PIE) may be quite important in the temperature range of $10^{3}$ - $10^{4}$ K, at which an upturn in $\Lambda (T)$ occurs, and photoionization heating can balance cooling. The gas can then expand at approximately constant temperature and continue to radiate (also see \cite{2003ApJ...590..791S}, \cite{2004ApJ...610..226S}). This, of course, is only possible if $\beta$ is low enough to even produce a transonic solution instead of radiating too strongly and producing a galactic fountain. The maximum $\beta$ that results in a transonic wind depends on the SFR, and a range of $\beta$ that can produce plausible winds matching the observed X-ray luminosity vs SFR relationship will be discussed in Sec. \ref{X-ray}. Of course, winds at temperatures below $10^{4} K$, at which point photoionization becomes important, do not radiate in X-rays. Thus, our treatment of gas at this low temperature has no impact on the agreement we find between the observed and calculated relationship between X-ray luminosity and SFR. It may be important for other emission diagnostics.

\newpage
\section{Comparing to X-Ray Luminosity Observations}
\label{X-ray}
Finally, we would like to test our model against recent observations and try to deduce a general relationship between mass-loading factor and SFR (or galaxy mass). Following the work of \cite{2014ApJ...784...93Z}, we calculate the theoretical X-ray luminosities, $L_{x}$ predicted by the CC model and predicted by our more general, modified model for a wide range of star formation rates. X-ray luminosities of actual star-forming galaxies are observed to have a linear correlation between $L_{x}$ and SFR (\cite{2014MNRAS.437.1698M}); however, \cite{2014ApJ...784...93Z} showed that the CC model gives a quadratic relationship $L_{x} \propto \text{SFR}^{2}$. This is because the CC solution predicts that $n \propto \text{SFR}$, and $L_{x} \propto n^{2}$. In addition, for a CC-like wind to have a plausible X-ray luminosity, the mass-loading factor $\beta = \frac{\dot{M}}{\text{SFR}}$ should be less than unity (\cite{2014ApJ...784...93Z}). Otherwise, the predicted X-ray luminosity is orders of magnitude greater than observed values. This is in contradiction to many galaxy formation codes, which require a mass-loading factor between $\beta = 1 - 10$ to reproduce the observed galaxy stellar mass function for low-mass galaxies. For example, \cite{2012MNRAS.422.2816B} compare many models of wind feedback, typically finding that high $\beta$ values give the best fit to the mass function; \cite{2013MNRAS.430.3213B} assume a constant $\beta = 2$ in their simulations; and \cite{2013MNRAS.428.2966P} find that $\beta = 2$ for a wind velocity of $484$ km/s reproduces the low-mass end of the galaxy stellar mass function quite well assuming that wind velocity decreases and mass-loading increases with decreasing galaxy mass.

The most recent observationally determined relationship between the total X-ray luminosity and SFR in star-forming galaxies is given by \cite{2014MNRAS.437.1698M}. In the 0.5 - 8 keV band, 
\begin{equation}
\frac{L_{x (0.5-8 \text{keV})}^{tot}}{\text{SFR}} \approx (4.0 \pm 0.4) \times 10^{39} \text{erg} s^{-1} / (\text{M}_{\odot} yr^{-1})
\end{equation}

Let's first consider a wind without energy input ($\alpha = 0$). We consider the same two transonic solutions from Fig. \ref{TwoSolsforLx} and re-scale the scaled velocity, sound speed, and density functions. We get the temperature $T = m c_{s}^{2}/(\gamma k_{b})$ where $k_{b}$ is Boltzmann's constant, $\gamma = 5/3$, and $m = \mu m_{H}$ is the mean mass per wind particle assuming the mean molecular weight $\mu = 0.61$ for solar abundances. 

The X-ray emission from our wind solutions is then 
\begin{equation}
L_{x (0.5-8 \text{keV})}^{\text{wind}} = \int n_{e}n_{H} \Lambda^{0.5-8 \text{keV}}(T,Z) dV
\end{equation}
The electron and hydrogen number densities, $n_{e}$ and $n_{H}$, respectively, are obtained from the wind density by $n(r) = \rho(r)/m$ and then $n_{H} = \chi n = 0.71 n$ for solar abundances. We use the XSPEC package (https://heasarc.gsfc.nasa.gov/xanadu/xspec/ ; Version 12.8.2) to calculate the cooling function $\Lambda^{0.5-8 \text{keV}}(T,Z)$ of a hot plasma in CIE, and in keeping with the XSPEC documentation, we assume $n_{e} = 1.2 n_{H}$. This ionization level is actually a function of temperature, but it is almost exactly 1.2 for all temperatures that contribute to emission in the 0.5-8 keV range. 

\begin{figure*}[!hb]
\label{varyBetaT0_onePlot}
\centering
\includegraphics[scale = 0.92,angle=0]{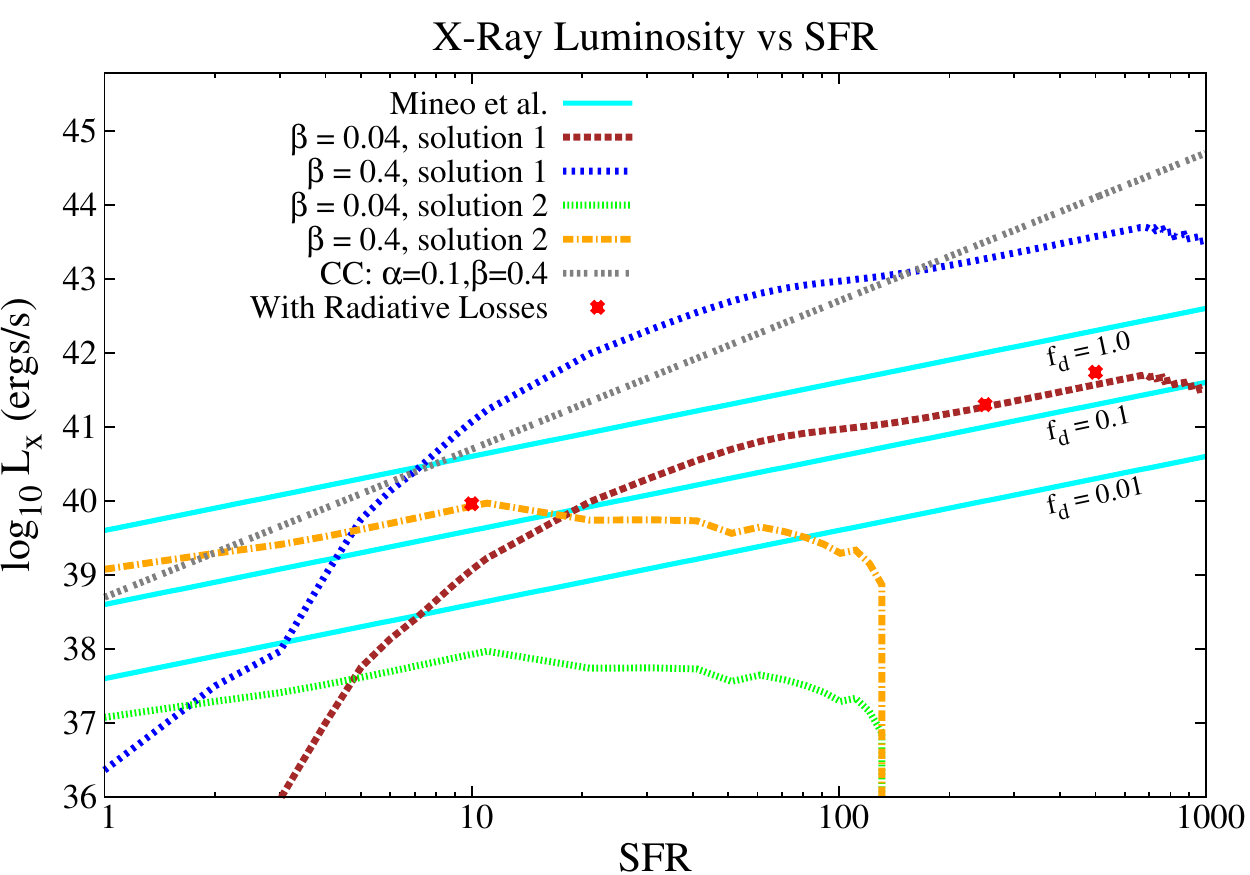}
\caption{Plot showing the wind X-ray luminosity vs SFR for our model and the observed trend of \cite{2014MNRAS.437.1698M} scaled by $f_{d}$ (blue lines). Results show that increasing the mass-loading factor $\beta$ shifts the luminosity upwards at all SFRs, and varying the temperature of the wind (solution 1 has lower temperature; solution 2 has higher temperature) shifts the region in which the $L_{x} - \text{SFR}$ relationship is linear. Using a fairly low central wind temperature, for low SFR galaxies, temperatures at many radii are too low to be picked up in the 0.5 - 8 keV band. Consequently, the low SFR region exhibits very low X-ray luminosities. If we choose a solution with higher central temperature, the low SFR galaxies produce a more linear relationship. If the solution curve has a central temperature great enough, the high SFR galaxies produce wind temperatures greater than 8 keV, causing the total X-ray luminosity in that band to fall off rapidly. The red stars show the effect of including radiative losses to points along the $\beta = 0.4$, solution 1 and the $\beta = 0.04$, solution 2 curves. The result is a marginal increase in luminosity.}
\end{figure*}

Only some fraction, $f_{d} \le 1$ of this diffuse emission will be from the hot wind. In M82, $f_{d} \approx 0.1$ (\cite{2009ApJ...697.2030S}); however, many of the galaxies included in Mineo's sample are not M82-like. In fact, as shown in our results in Fig. \ref{varyBetaT0_onePlot}, including a linear relationship between galaxy mass and SFR, as sketched in Fig. 1 of \cite{2014MNRAS.437.1698M}, the predicted $L_{x} - \text{SFR}$ relationship is no longer strictly quadratic as was predicted by the CC model. The linear galaxy mass to SFR relationship we used in Fig. \ref{varyBetaT0_onePlot} was

\begin{equation}
\label{M_SFR}
M = 10^{10}  \times 0.1 \times \text{SFR} (\text{M}_{\odot})
\end{equation}
which is a rough fit to the sample in \cite{2014MNRAS.437.1698M}. It should be noted that the \cite{2014MNRAS.437.1698M} sample excludes many starburst galaxies that can have low mass but high SFRs and may lie on a relationship similar to eqn. (\ref{M_SFR}) but with a different constant of proportionality. We have found that assuming a linear relationship between SFR and galaxy mass is key to reproducing the X-ray luminosity vs SFR relationship observed by \cite{2014MNRAS.437.1698M}. 

Even without energy input ($\alpha = 0$), Fig. \ref{varyBetaT0_onePlot} shows a linear correlation between $L_{x}$ and SFR for various ranges of SFR depending on which transonic solution we choose from Fig. \ref{TwoSolsforLx}. Using a fairly low central wind temperature, for low SFR galaxies, temperatures at many radii are too low to be picked up in the 0.5 - 8 keV band. Consequently, the low SFR region exhibits very low X-ray luminosities. If we choose a solution with higher central temperature, the low SFR galaxies produce a more linear relationship. If the solution curve has a central temperature great enough, the high SFR galaxies produce wind temperatures outside the 8 keV limit, causing the total X-ray luminosity in that band to fall off rapidly. Therefore, the galactic central conditions can play a large role in what the $L_{x} - \text{SFR}$ correlation is in this 0.5 - 8 keV band. 

As seen in Fig. \ref{varyBetaT0_onePlot}, if we use a higher mass-loading factor $\beta$, the X-ray luminosity shifts upwards for each SFR because the wind density increases; however, radiative losses can also be important for all $\beta$. The red stars show the effect of including radiative losses to points along the $\beta = 0.4$, solution 2 and the $\beta = 0.04$, solution 1 curves. It should be noted, however, that in the range of $\text{SFR} \approx 20-100 \text{M}_{\odot}/yr$, the central temperatures of the non-radiative $\beta = 0.04$, solution 1 curve cannot drive a radiative wind with $\beta = 0.04$. At this range of SFR, $\dot{M} = \beta \text{SFR}$ is high enough, even with a low $\beta$, that higher temperatures are required to drive a radiative wind. Therefore, radiative wind luminosities are not compared to non-radiative wind luminosities in that SFR range. 

Generally, radiative losses increase the X-ray luminosity of the wind in the temperature range of $10^{5} - 10^{7}$ K. In this range, the emissivity is greater for lower temperatures. As seen in Fig.\ref{radiative_varyBeta}, including radiative losses gives the wind an overall lower temperature than for the non-radiative wind, meaning the emissivity can be factors of ten greater depending on how significant the radiative losses are. Also, as the wind propagates outwards from $r \approx 0$, some thermal energy that would have otherwise been converted to kinetic energy is now lost to radiation. Consequently, the wind doesn't spread outward as quickly and, instead, clumps up, resulting in an increase in density near the galactic center. This increase in density and decrease in temperature both serve to enhance the radiative loss, and hence increase the X-ray luminosity. The resulting increase is fairly marginal, though, and doesn't change the linear $L_{x} - \text{SFR}$ behavior. 

\begin{figure*}[!ht]
\label{varyBetaT0}
\centering
\includegraphics[scale = 0.45, angle = -90]{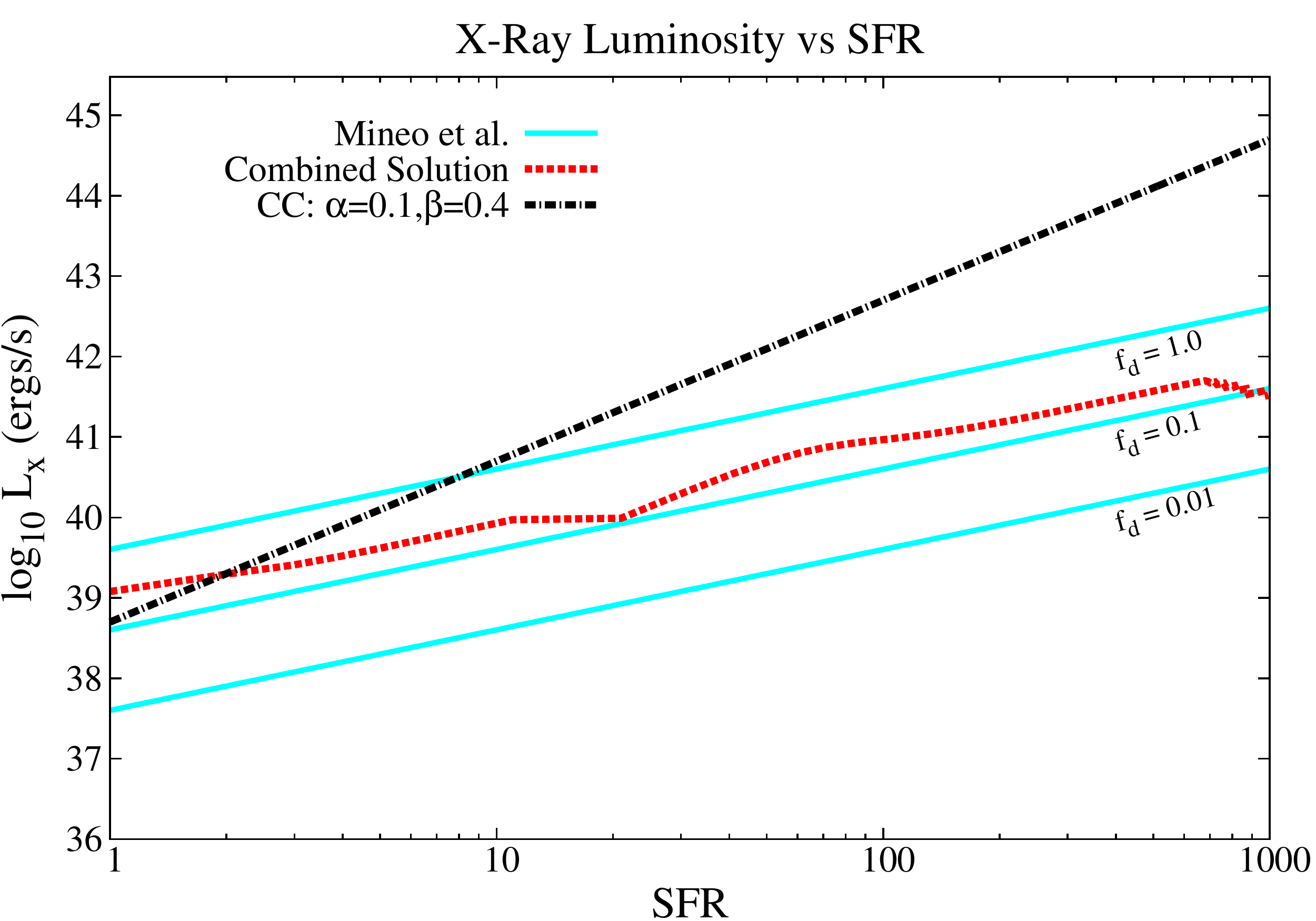}
\caption{Using solution 1 and $\beta = 0.04$ for $\text{SFR} > 20$, solution 2 and $\beta = 0.4$ for $\text{SFR} < 20$. By increasing the mass-loading factor and initial scaled wind temperature for low-mass galaxies, we can roughly create a linear relationship between X-ray luminosity and SFR. Since we do not have an exact value for $f_{d}$ for each galaxy in the sample, we cannot use this plot to constrain $\beta$; however, the possible relationship that scaled velocity and mass-loading efficiency should be higher for low-mass galaxies gains some validation from this plot in that the expected linear correlation between X-ray luminosity and SFR is at least reproduced. As shown in \cite{2013MNRAS.428.2966P}, this increase in $\beta$ for low-mass galaxies can help reproduce the observed galaxy stellar mass function in simulations.}
\end{figure*}

It has been observed (\cite{2014arXiv1412.2139C}) that higher mass galaxies (and hence higher SFR galaxies according to the \cite{2014MNRAS.437.1698M} trend) have lower mass-loading factors. This has also been seen in hydrodynamical simulations of galactic winds (\cite{2012MNRAS.421.3488H}). Using $\beta = 0.4$ for low-mass ($\text{SFR} < 20  \text{M}_{\odot}\text{/yr}$) galaxies and $\beta = 0.04$ for high-mass ($\text{SFR} > 20  \text{M}_{\odot}\text{/yr}$) galaxies, coupled with using a solution curve with higher scaled initial temperature for low-mass galaxies, we can produce a linear correlation for almost the entire range of SFRs (see Fig. \ref{varyBetaT0}). 

It should be stressed that we have shown just one such combination of mass-loading factor, SFR, and wind solution that produces a linear $L_{x} - \text{SFR}$ relationship. Many such combinations exist. It is important, though, that one cannot produce a linear $L_{x} - \text{SFR}$ relationship across a large range of SFR without switching the wind solution and/or $\beta$ for various ranges of SFR. 

\begin{figure}[!ht]
\label{max_min_Beta_SFR}
\centering
\hspace*{-0.5 cm}\includegraphics[scale = 0.28]{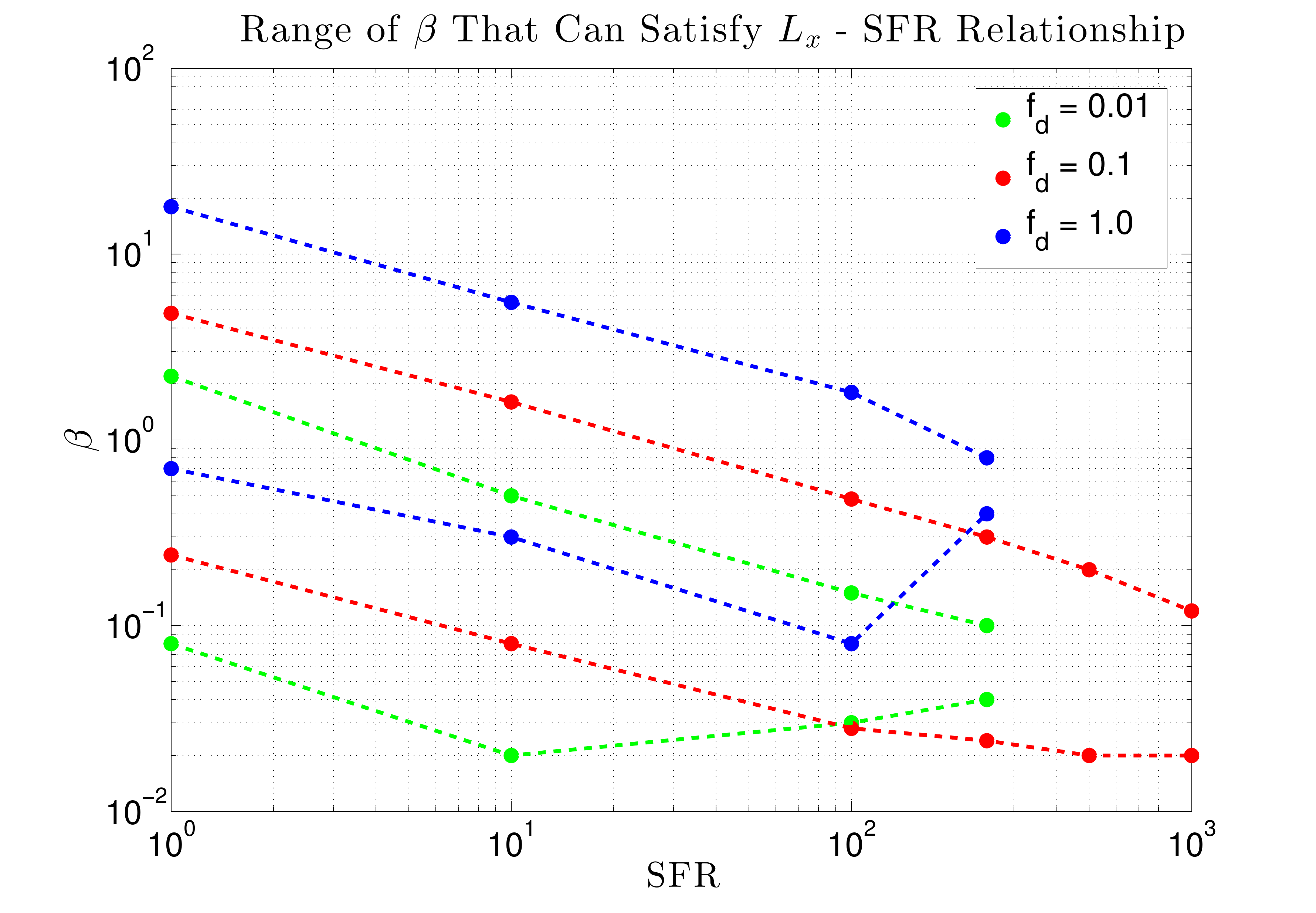}
\caption{Assuming $f_{d} = 0.01$, $0.1$, and $1.0$ and a maximum central wind temperature of $5 \times 10^{8}$ K, these are the minimum and maximum $\beta$ that can reproduce the target X-ray luminosity at each SFR following the trend in the 0.5-8 keV range from \cite{2014MNRAS.437.1698M} scaled by $f_{d}$. Note that for SFR $\agt 250$, the $f_{d} = 0.01$ and $f_{d} = 1.0$ luminosities cannot be reproduced by our model. Overall, there is a clear inverse relationship between allowed $\beta$ and SFR, and hence between $\beta$ and galaxy mass, as we assume a direct relationship between SFR and galaxy mass.}
\end{figure}

In fact, we find that only certain values of $\beta$ are allowed if we would like to construct a linear $L_{x} - \text{SFR}$ relationship from radiative outflows while keeping the central galactic temperature to a reasonable value. In Fig. \ref{max_min_Beta_SFR}, we look at the space of mass-loading values that can fit the observed X-ray luminosity trend scaled by $f_{d} = 0.01, 0.1, \text{and} 1.0$. For each outflow, as shown in Fig. \ref{centralTemp}, the central temperature must be above a certain value dependent on the mass-loading factor for that wind. This sets a lower bound on $\beta$ because if a wind is not mass-loaded enough, it can only radiate brightly if the temperature is near the peak of the cooling curve. These low central temperatures cannot always be achieved, though. Similarly, we can set an upper bound on $\beta$. For highly mass-loaded winds, the temperature must be high, i.e. away from the cooling curve peak, to limit the luminosity. At some point, we consider this central galactic temperature requirement to be unreasonably high. We choose this temperature to be $T_{0}^{\text{max}} = 5 \times 10^{8} K$, so only winds with central temperatures below this value are considered. 

For higher SFR, more mass is being injected into the wind for the same $\beta$; consequently, as seen in Fig. \ref{max_min_Beta_SFR}, there must be a decreasing trend between $\beta_{min}$ and $\beta_{\text{max}}$ versus SFR, and hence versus galaxy mass, to limit the luminosity and keep the central temperature under the $5 \times 10^{8}$ K limit. As shown in \cite{2013MNRAS.428.2966P}, increasing $\beta$ for low-mass galaxies helps reproduce the observed galaxy stellar mass function in their simulations. Therefore, we believe that the positive effect this scaling has on fitting the expected $L_{x} - \text{SFR}$ relationship provides further validation, along with observations, to use this assumption in future galaxy formation simulations.  

For higher luminosities (for greater $f_{d}$), the range of $\beta$ shifts upward, as a denser wind will radiate more strongly; however, it is interesting that, for a high $f_{d} = 1.0$ and a low $f_{d} = 0.01$, the range of $\beta$ gets squeezed at high SFR until, at SFRs $\agt 250$,  the target luminosities can no longer be achieved by our model for any value of $\beta$. The $f_{d} = 0.1$ trend, however, can be reproduced for the entire range of SFR, albeit with the range of $\beta$ again being squeezed at high SFRs. This squeezing occurs because, for high SFR, the luminosity trends higher and higher; therefore, if $\beta$ is too low, the wind will not radiate to the necessary extent. This means the slope of the minimum $\beta$ curve must become less negative, all while the maximum $\beta$ curve, which is constrained by the $5 \times 10^{8}$ K limit on central wind temperature, continues to decrease with roughly the same slope. 

\clearpage

\section{Conclusions}
We have re-worked the \cite{1985Natur.317...44C} model to include non-uniform mass and energy source distributions, the gravitational potential for an extended mass distribution, and radiative losses. After scaling our steady-state equations, we generate many different transonic wind solutions, each for a different central wind temperature, for an infinite number of galactic masses and radii. Using this set-up, we can easily explore the space of central temperature, mass-loading factors, and energy-loading factors.

In Fig. \ref{tcool}, we show the ratio of the cooling timescale to the dynamical timescale of our wind model for various star formation rates and mass-loading efficiencies. This provides a first order estimate as to whether radiative losses should be included. The cooling timescale trends lower for lower SFR and also for higher $\beta$; however, we also see that radiative cooling is not as important if we use a scaled solution with higher initial temperature.

When we include radiative loss physics, we find that generating a transonic solution is more expensive for more heavily mass-loaded winds, i.e. the wind must have a higher central temperature. For these high - $\beta$ transonic winds, then, the cooling time is generally greater, hence the radiative losses incurred by increasing density are somewhat lessened by the higher temperature required to get the transonic solution in the first place. Conversely, low-$\beta$ winds do not require as much thermal energy to drive them, i.e. they can be driven at lower temperatures closer to the peak of the cooling curve. Therefore, radiative losses can be important for low mass-loaded winds, as well, despite having a lower density. When radiative losses are significant, we find that including radiation actually increases the X-ray luminosity compared to a non-radiative wind with the same central temperature. 

With radiative losses included in our model, we also consider the efficiency of various transonic solutions, namely which $\alpha$ and $\beta$ parameters give rise to transonic solutions with the greatest energy flux per mass out of the galaxy. We find that, for high $\beta$, radiative losses are very important and decrease the asymptotic temperature and velocity of the wind. However, efficiency can be recovered if more energy is injected. 

Also, given the infinite number of possible subsonic solutions, we consider how efficient subsonic winds can be in expelling mass from a galaxy. We find that, due to radiative losses again, these winds lose their energy per mass very quickly as the wind cools towards temperatures near the peak of the cooling curve. We define the ``wind extent" as how far the subsonic solution propagates before this large drop in $\epsilon$, and we find that winds with higher energy injection can extend further than other subsonic solutions but still only so far. Transonic solutions, on the other hand, can extend to infinite distances in our model, albeit with a density that decreases as $1/r^2$ outside the galaxy radius. Therefore, we conclude that transonic solutions, especially those with low $\beta$ and high $\alpha$, are the most efficient in driving mass from a galaxy. These low-$\beta$ radiative solutions are also the least expensive to drive in that they have the lowest central temperature requirement. 

We can also generate many transonic wind solutions that decrease in temperature very rapidly due to radiative cooling, possibly explaining the existence of cool, fast outflows from galaxies. One example for a relatively low $\beta = 0.04$ is given in Fig. \ref{lowTempFig}. 

We also compare the predicted X-ray luminosity of various outflows to recent $0.5 - 8$ keV observations for a wide range of SFRs. This provides a first observational test for our model compared to the CC model. We find that, when the wind temperature at all radii gives energies between 0.5 - 8 keV, i.e. when the central wind conditions are right to do so, our modified CC model with a non-uniform mass source and extended mass distribution produces a linear correlation between total X-ray luminosity and SFR for various ranges of SFR. Choosing a higher central temperature and higher mass-loading factor for low-mass galaxies, we can produce this linear correlation over the entire range of SFR from $1 - 10^{3} \text{M}_{\odot}\text{/yr}$. We present only one combination of mass-loading and wind solution resulting in a linear $L_{x}-\text{SFR}$ relationship; however, many combinations can achieve this, as long as mass-loading and solution are not fixed over the entire range of SFR. Specifically, we find that, assuming various scaling factors $f_{d}$, there is a range of $\beta$ for each SFR that can fit the observed luminosity. The allowed $\beta$ values typically decrease as SFR, and hence galaxy mass, increases. This is supported by recent observations that find an inverse relationship between mass-loading and galaxy mass (\cite{2014arXiv1412.2139C}), and it lends support to many galaxy formation simulations which utilize higher mass-loading factors for low-mass galaxies to help reproduce the observed stellar mass function at the low-mass end. 

\clearpage
\acknowledgments{
We would like to thank the referee for very insightful comments, as well as Jay Gallagher and John Chisholm for many helpful discussions on wind observations and low-temperature winds. We also thank Dan McCammon for helping us obtain cooling curves from the XSPEC package and Chris Bard for insights on determining critical points. This work is  funded by NSF Grant No AST-1211258 and ATP NASA Grant No NNX144AP53G. We also acknowledge support from the WARF Foundation and the College of Letters and Science at the University of Wisconsin-Madison. ED gratefully acknowledges the support of the Alfred P. Sloan Foundation and expresses appreciation towards the Aspen Center for Physics for their hospitality, funded by the NSF under Grant No. PHYS-1066293.}

\renewcommand{\theequation}{A\arabic{equation}}
\setcounter{equation}{0}  
\section*{Appendix}

\label{appendix}

The solution to the continuity equation with spherical symmetry, eqn. (\ref{massEQN}), is
\begin{equation}\label{rhovar}
\rho(r)=\frac{1}{r^2u}\int_{0}^{r}s^2q(s)ds
\end{equation}
which can also be written
\begin{equation}\label{rho2}
\rho=\frac{r\langle q\rangle}{3u},
\end{equation}
where
\begin{equation}\label{meanq}
\langle q\rangle\equiv\frac{3}{r^3}\int_{0}^{r}s^2q(s)ds
\end{equation}
is the mean $q$ inside radius $r$.

The outward force is provided by the pressure gradient (here let's just consider thermal gas pressure). For an ideal gas equation of state
\begin{equation}
\frac{dP}{dr}=\frac{k_BT}{m}\frac{d\rho}{dr}+\frac{\rho k_B}{m}\frac{dT}{dr}.
\end{equation}
The temperature is determined by the first law of thermodynamics; $dU=dQ+Pd\rho/\rho^2$ (where $dQ=TdS$ and $Q$ is not to be confused with the CC source term in their energy equation). Using $P=(\gamma - 1)\rho U$ and assuming a steady state with radial flow $u$, the temperature equation is given by eqn. (\ref{myTEQN}), and the pressure described by eqn. (\ref{myPEQN}).

The main equation that is integrated numerically is the steady state radial momentum equation, eqn. (\ref{momEQN}).

Substituting the expression for dp/dr and plugging in the result of the continuity equation, our main velocity equation, without radiative losses included, is
\begin{multline}
     \rho u \frac{\partial u}{\partial r}(1- \frac{c_{s}^{2}}{u^{2}}) =  -qu(1+ \frac{c_{s}^{2}}{u^{2}}) - \frac{(\gamma - 1)}{u} \rho \frac{\partial Q}{\partial t} \\
      - \rho \frac{\partial \Phi}{\partial r} + \frac{2c_{s}^{2}<q>}{3u}
\end{multline}

It is logical to scale the above equation by $\frac{GM}{R}$.
We also scale r by R, i.e. $\tilde{r} = r/R$, $\tilde{q}(\tilde{r}) = q_{0}(1-\tilde{r}^{2})$ and $<\tilde{q}(\tilde{r})> = q_{0}(1-3/5 \tilde{r}^{2})$. 

To add energy sources to our model, we define an energy per mass function $\eta$ such that  $\rho \frac{\partial Q}{\partial t} = \eta q$. Scaling by $\frac{GM}{R}$ means $\tilde{\eta} = \eta /\frac{GM}{R}$.

Then, $\tilde{u} = u/\sqrt{\frac{GM}{R}}$, and our continuity equation, which was $\frac{1}{r^{2}} \frac{\partial{(\rho u r^{2})}}{\partial r} = q(r)$, becomes 

\begin{equation}
\frac{1}{\tilde{r}^{2}} \frac{\partial{(\rho \tilde{u} \tilde{r}^{2})}}{\partial \tilde{r}} = \tilde{q}(\tilde{r}) \sqrt{R^{3}/GM} 
\end{equation}

Here, $\rho = \frac{<q>r}{3u}= \frac{<\tilde{q}>\tilde{r}}{3\tilde{u}} \sqrt{\frac{R^{3}}{GM}}$

The pressure equation is similarly changed. $\frac{\partial P}{\partial r} = c_{s}^{2} \frac{\partial \rho}{\partial r} + \frac{(\gamma-1)\rho}{u} \frac{\partial Q}{\partial t}$ becomes
 
\begin{equation}
 \frac{\partial \tilde{P}}{\partial \tilde{r}} = \tilde{c}_{s}^{2} \frac{\partial \tilde{\rho}}{\partial \tilde{r}} + \frac{(\gamma-1)}{\tilde{u}} \sqrt{\frac{R}{GM}} \sqrt{\frac{GM}{R}} \tilde{\eta} \tilde{q}
 \end{equation}

Using this scaling, our main non-radiative equation with scaled variables becomes

\begin{multline}
\label{radU}
     \frac{<\tilde{q}>\tilde{r}}{3} \frac{\partial \tilde{u}}{\partial \tilde{r}}(1- \frac{\tilde{c}_{s}^{2}}{\tilde{u}^{2}}) =  -\frac{<\tilde{q}>}{3\tilde{u}}(\tilde{r}^{2}+ \tilde{c}_{s}^{2}) \\
      - \tilde{q}\tilde{u}
      - \frac{\tilde{c}_{s}^{2}}{\tilde{u}}(\tilde{q} - <\tilde{q}>) - \frac{(\gamma - 1)}{\tilde{u}} \tilde{\eta} \tilde{q}
\end{multline} 

Similarly, the scaled temperature equation is
\begin{multline}
\label{radCS}
	\frac{\partial \tilde{c}_{s}^{2}}{\partial \tilde{r}} = \frac{3(\gamma-1)\gamma \tilde{\eta} \tilde{q}}{<\tilde{q}>\tilde{r}} \\
	 + \frac{3(\gamma-1) \tilde{c}_{s}^{2} \tilde{u}}{<\tilde{q}>\tilde{r}} \Big[ \frac{-2<\tilde{q}>}{3\tilde{u}}-\frac{<\tilde{q}>\tilde{r}}{3\tilde{u}^{2}} \frac{\partial \tilde{u}}{\partial \tilde{r}} + \frac{\tilde{q}}{\tilde{u}} \Big]
\end{multline}
 
The process of scaling by $\frac{GM}{R}$ is similar when radiative losses are included; however, since the cooling depends on temperature, we cannot completely scale $\frac{GM}{R}$ out of the radiative loss terms, meaning that we must specify the quantity $\frac{GM}{R}$ before we generate a wind solution. For non-radiative winds, $\frac{GM}{R}$ could be cancelled out of each term in the scaled momentum equation, allowing us to generate a scaled solution and then specify $\frac{GM}{R}$ after the fact. 

\clearpage
\bibliography{winds_apj_2015_revised}

\end{document}